\documentclass{aawithoutcopy}
\usepackage{graphicx}
\usepackage{psfig}
\usepackage{txfonts}
\usepackage{rotating}
\usepackage{amssymb}
\begin{document}
\title{Quasar Host Galaxies in the FORS Deep Field}
\subtitle{}
\author{C. Villforth\inst{1,2,3} \and J. Heidt\inst{3} \and K. Nilsson\inst{1}}

\institute{Tuorla Observatory, University of Turku, V\"{a}is\"{a}l\"{a}ntie 20, 21500 Piikki\"{o}, Finland\\
\and
Nordic Optical Telescope, Apartado 474, 38700 Santa Cruz de La Palma, Spain\\
\and
ZAH, Landessternwarte Heidelberg, K\"{o}nigstuhl 12, 69117 Heidelberg, Germany
}

   \date{Received 04.03.2008; Accepted 02.07.2008} 
\abstract
{The evolution of quasar host galaxies is still hardly studied at high redshifts ($z>2$), although this is a very interesting redshift range as both the quasar activity and the star formation rate density have their peak at $z\approx2$--$3$. This makes is especially interesting to study properties of quasar host galaxies, such as the star formation rate or the black hole mass at this redshift. A proper classification of quasar host galaxies at high redshift would help to answer the question which role quasars play in galaxy evolution.}
{In this paper we study different properties of quasars and their host galaxies at high redshifts up to $z\approx3.4$. We compare our results to those of other authors and discuss the correlation between galaxy evolution and quasar activity.}
{We analysed broad-band images in eight filters (from U to K) of eight quasars in the FORS Deep Field with redshifts between $z=0.87$ and $z=3.37$. A fully 2-dimensional decomposition was carried out to detect and resolve the host galaxies. Using the magnitudes in different filters, we investigated the presumed galaxy type, galaxy age, star formation rate and the mass of the central black hole of the host galaxies. In addition, the masses of the central black hole for the whole sample were derived using the corresponding quasar spectra.}
{We were able to resolve the host galaxies of two out of eight quasars between $z=0.87$ and $z=2.75$. Additionally, two host galaxies were possibly resolved.}
{The resolved low-redshift quasar ($z=0.865$) was identified as a late type galaxy with a moderate star formation rate of $1.8 M_{\odot}/yr$ hosting a supermassive black hole with a mass of $\lesssim10^{8}M_{\odot}$. The resolved high redshift host galaxy ($z=2.7515$) shows moderate star formation of $4.4$--$6.9 M_{\odot}/yr$, for the black hole mass we found a lower limit of $>10^{7}M_{\odot}$. All quasars host supermassive black hole with masses in the range $\sim10^{7}$--$10^{9}M_{\odot}$. Our findings are well consistent with those of other authors.}

\maketitle

\section{Introduction}

The quasar activity and the star formation rate density apparently peak at $z\approx2$--$3$ (see e.g. Nandra et al. \cite{Na05}, Appenzeller et al. \cite{Ap04}). This implies a strong link between quasar activity and galaxy formation. Additionally, this is the redshift range at which the luminosity of quasar host galaxies seems to turn over (Falomo et al. \cite{Fa08}), even though this turnover is poorly constrained due to a small number of resolved host galaxies at high redshifts. Although the formation and evolution of nuclear activity is not yet understood it seems to be clear that there is a strong link between the evolution of galaxies and nuclear activity. After the observational confirmation of supermassive black holes in nearby early-type galaxies, correlations between the properties of the host galaxy and its central black hole were found (e.g. Novak et al. \cite{No06}). This affirmed the assumption that the evolution of early-type galaxies and the central black hole, and thus nuclear activity, are linked.

The evolution of host galaxies of quasars has been studied over a wide redshift range from very low (e.g. Dunlop et al. \cite{Du03}; Percival et al. \cite{Pe01}) up to very high redshifts of $z\sim3$ (e.g. Ridgway et al. \cite{Ri01}; Falomo et al. \cite{Fa04}, \cite{Fa08}; Jahnke et al. \cite{Ja04}; Kotilainen et al. \cite{Ko07}). It is nowadays established that luminous quasars typically reside in luminous massive early-type galaxies. The luminosity of host galaxies seems to increase with redshift, with the host galaxies of radio-quiet quasars apparently being on average less luminous than the ones of radio-loud quasars (e.g. Kukula et al. \cite{Ku01}). Quasar host galaxies at high redshifts of $z=2\sim3$ have not yet been studied elaborately, since due to high QSO to host light ratios, extremely deep high resolution images are required to assure a sufficiently high signal to noise ratio. Consequently, most studies of quasar hosts were carried out using the Hubble Space Telescope (HST) (e.g. Kukula et al. \cite{Ku01}; Ridgway et al. \cite{Ri01}; Jahnke et al. \cite{Ja04}). Jahnke et al. (\cite{Ja04}) found rest-frame near-UV absolute magnitudes of --$20.0<M_{200nm}<$--$22.2$, but did not study morphological properties. Kukula et al. (\cite{Ku01}) found absolute V-band magnitudes around --$22<M_{V}<$--$24$, early-type morphologies and differing evolution of the host galaxies of radio-quiet quasars (RQQ) and radio-loud quasars. In their samples, host galaxies of RQQ are fainter than those of RLQ on average with a growing gap towards higher redshifts. Adaptive Optics (AO) observations are beginning to play a major role as well. Recent studies of high redshift quasars (Kotilainen et al. \cite{Ko07}, Falomo et al. \cite{Fa08}) using AO showed no deviation from former findings.

Since quasar host galaxy studies are typically carried out in 1--2 filters, galaxy types are derived via their morphological appearance. Thus, investigations of stellar populations, star formation rates or ages of the host galaxy are difficult or even impossible. Recently, a spectroscopic study of quasar host galaxies at low redshift ($z\sim0.3$) was performed by Letawe et al. (\cite{Le06}). Most host galaxies were found to have Sc-like populations, showing large amounts of ionised gas and signs of interaction. Canalizo \& Stockton (\cite{CS01}) analysed a sample of low-redshift radio-quiet quasars ($z\leqslant0.4$) spectroscopically. They found tidal interaction-induced starbursts in almost all of their objects. For 5/9 sources they found star-forming regions in the centre which is consistent with merger-induced starbursts. They found clues that central starbursts and quasar activity were triggered simultaneously. Hyv\"{o}nen et al. (\cite{Hy07}) analysed the colours of 18 low redshift ($z\leqslant0.3$) BL Lac objects and found old evolved stellar populations with an additional about 1 Gyr old young stellar population. Absence of signs of interaction implies that star formation and nuclear activity where not triggered simultaneously. Schramm et al. (\cite{Sc08}) studied high redshift ($z\approx3$) high luminosity quasars in restframe optical and found host with stellar populations indicating recent massive star formation. The different results seem to indicate that stellar populations of quasar host galaxies depend on the luminosity of the quasar itself, the radio-loudness and the redshift.

In this paper we present the analysis of eight quasars in the FORS Deep Field (see Heidt et al. \cite{heifdf}) with a redshift range from $z\approx0.9$-$3.4$. The field was observed with the VLT and HST from near UV to near IR (UBgRIJKs,F814W), allowing multiband analysis of the host galaxies from rest-frame UV to restframe optical. Due to very long exposure times (up to 44400s in U), the images are extremely deep in the optical with 50\% completeness limits up to 27.7 in the B-band. Additionally, the seeing is excellent, with the best seeing being 0.''53 in the I-band. This gives us a unique possibility to study the probable galaxy type, star formation rate and mass of central black hole of the host galaxies.

The paper is organised as follows. In Section 2 we shortly present the quasar sample and summarise the observations. In Section 3 we describe the data analysis. In Section 4 we discuss the results. We investigate the galaxy types, host galaxy formation timescales, star formation rates and masses of the central black holes and compare our results to those of other authors. We discuss our results regarding the link between galaxy formation and quasar activity in Section 5 and compare our results to those of other authors. Our results are summarised in Section 6.

All magnitudes are given in the Vega-system. The cosmology used is $H_{0}=70\textrm{km/s/Mpc}, \Omega_{\Lambda}=0.7, \Omega_{m}=0.3$, unless otherwise stated.

\section{The Quasar Sample and available Data}

\begin{figure*}[t]
\centerline{\hbox{
\psfig{figure=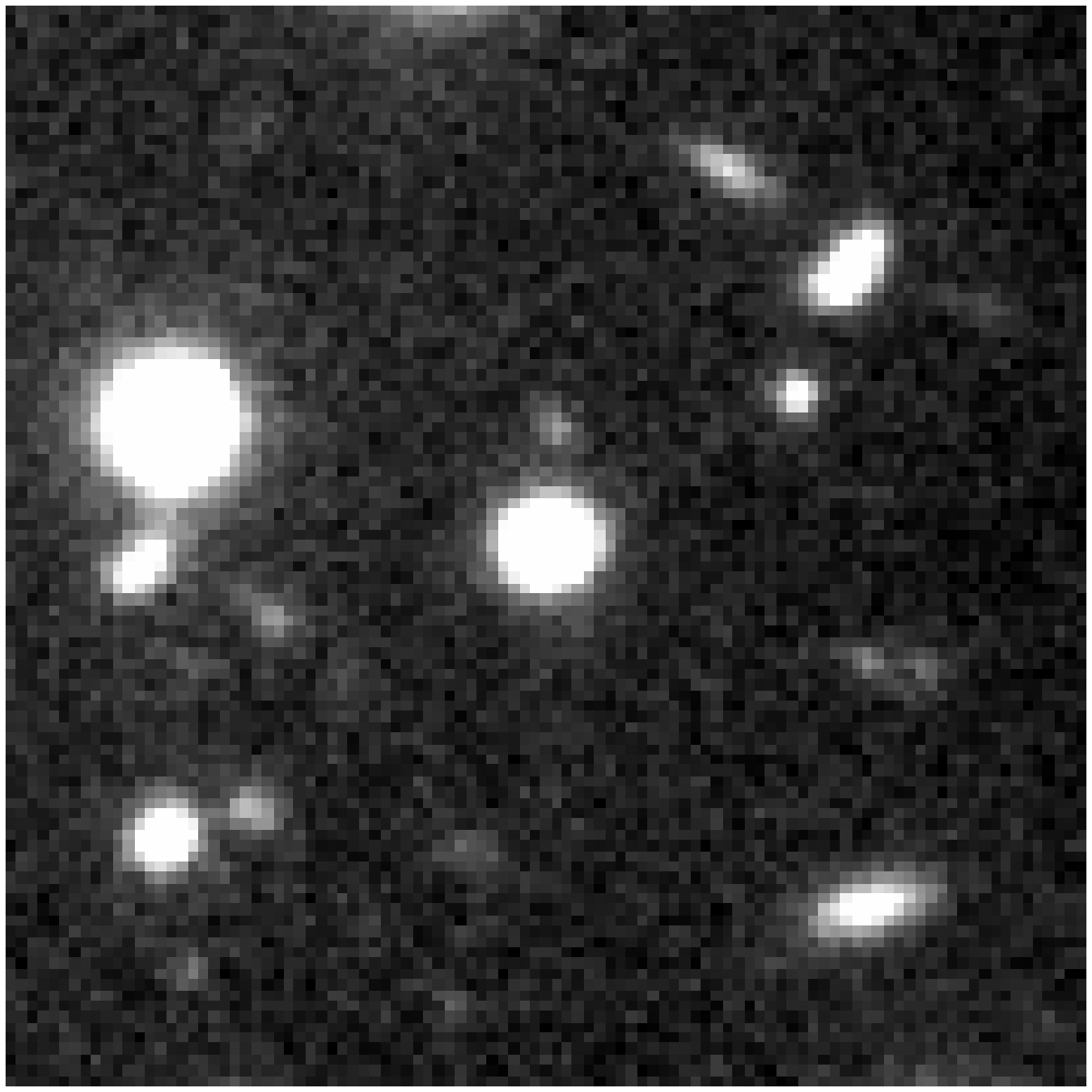,width=3.5cm,clip=t}
\hspace*{0.1cm}
\psfig{figure=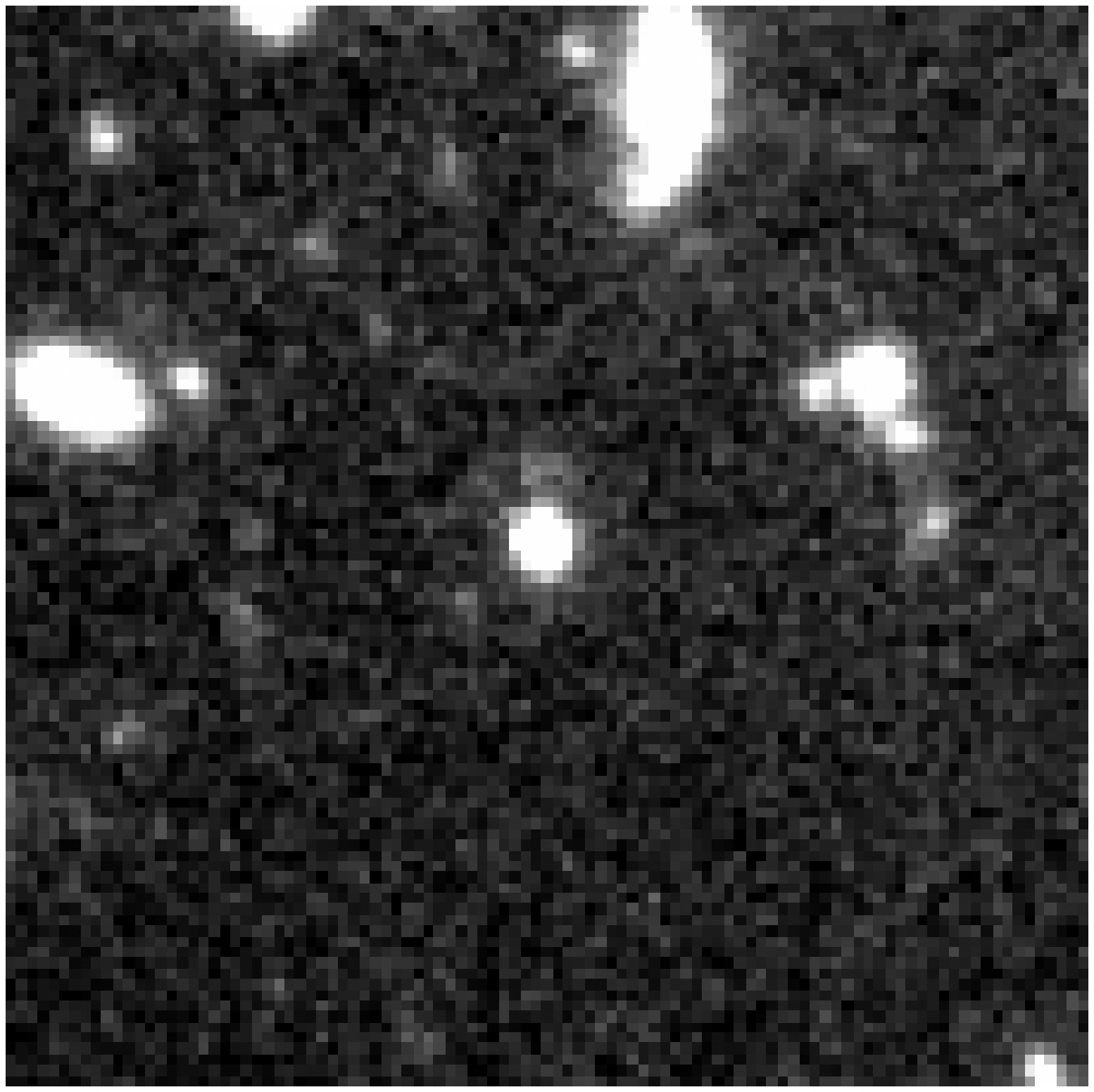,width=3.5cm,clip=t}
\hspace*{0.1cm}
\psfig{figure=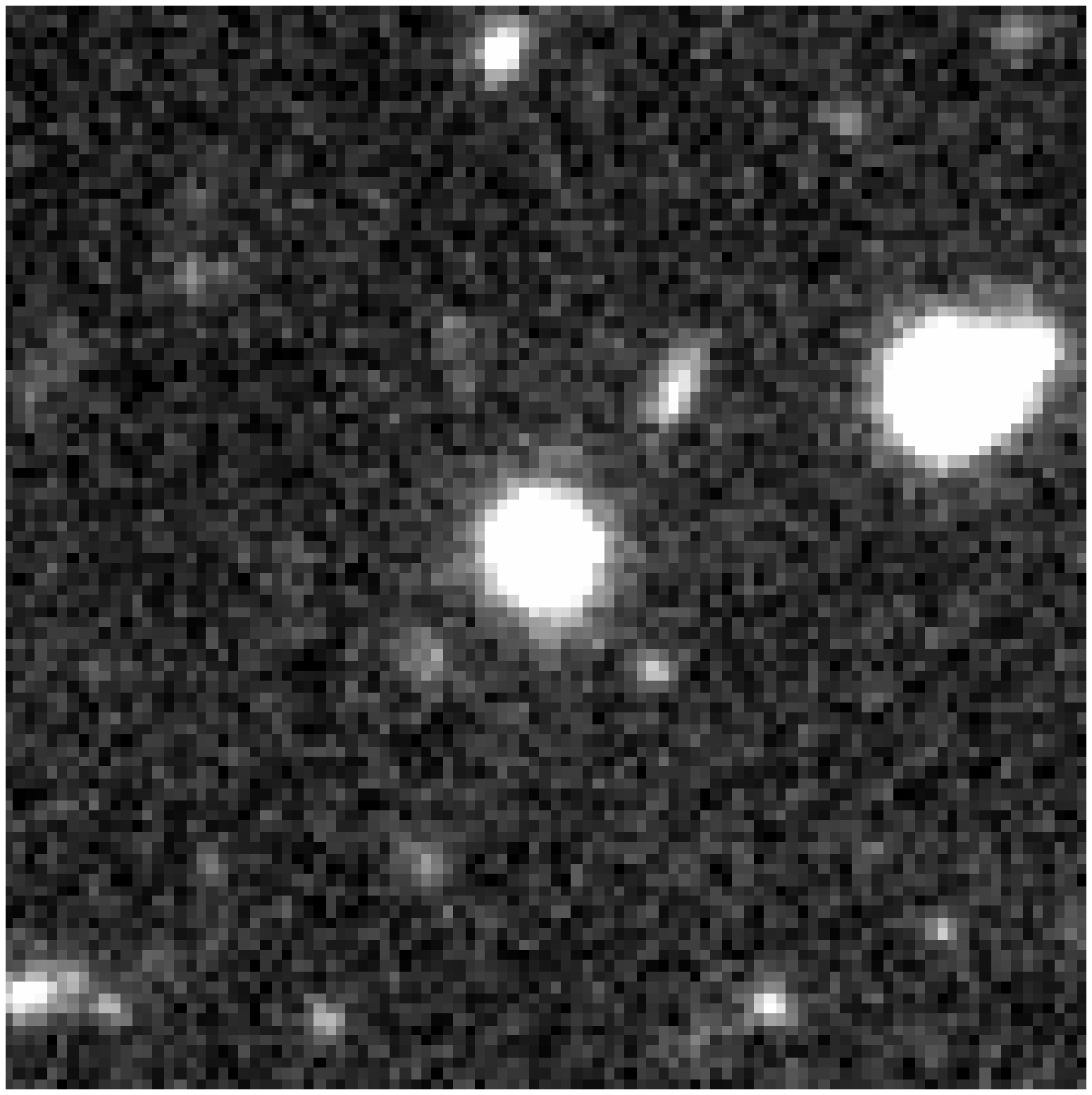,width=3.5cm,clip=t}
\hspace*{0.1cm}
\psfig{figure=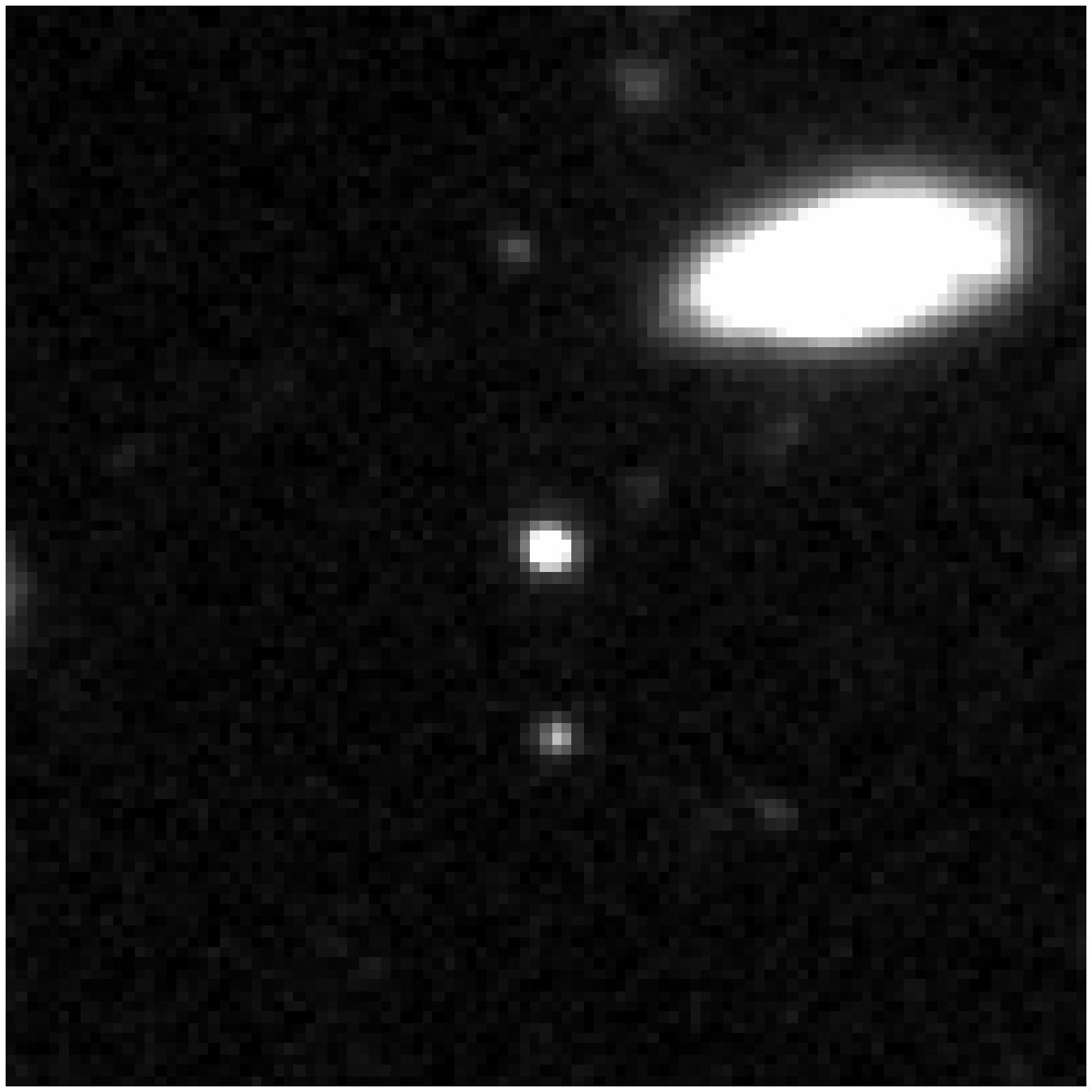,width=3.5cm,clip=t}
}}
\vspace*{1cm}
\centerline{\hbox{
\psfig{figure=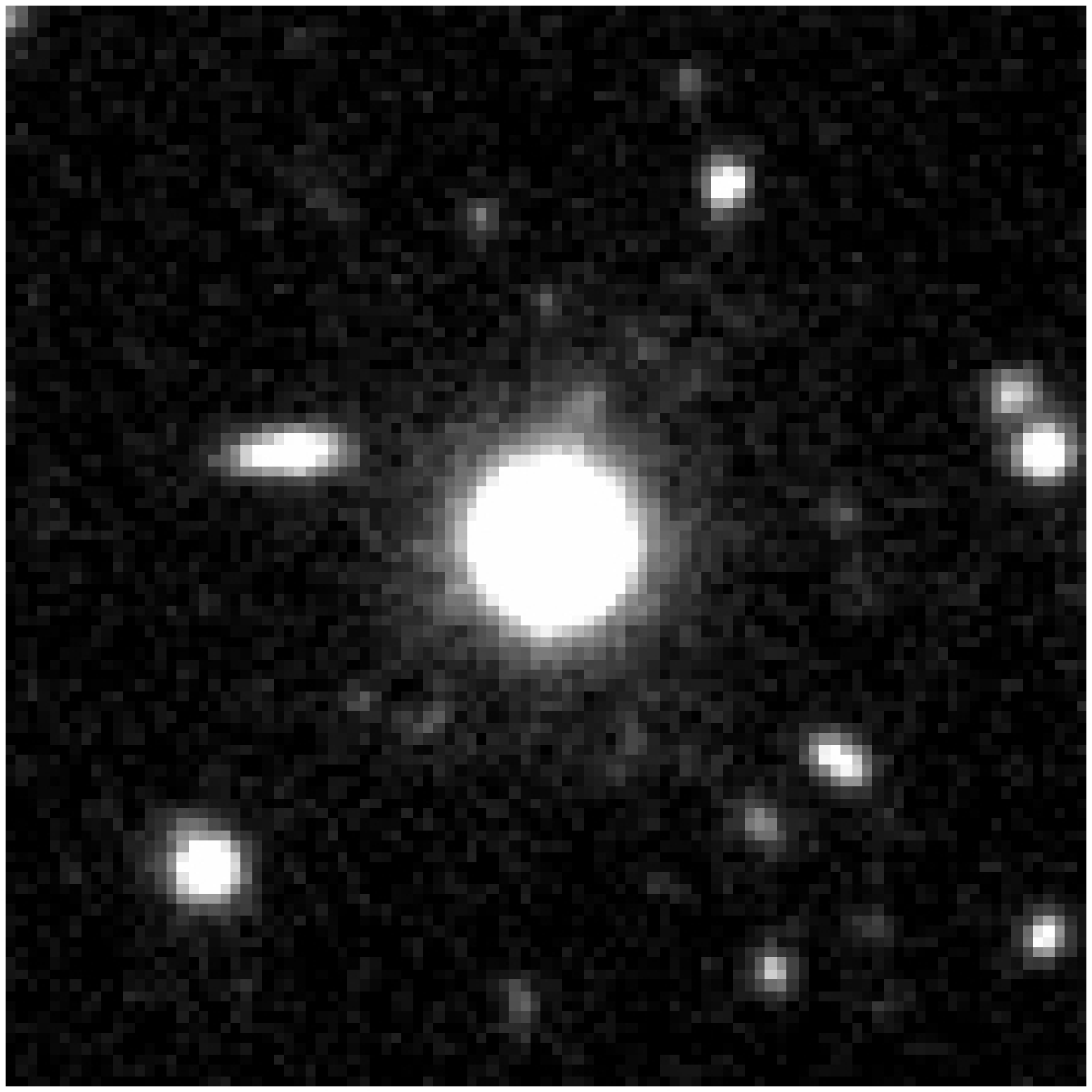,width=3.5cm,clip=t}
\hspace*{0.1cm}
\psfig{figure=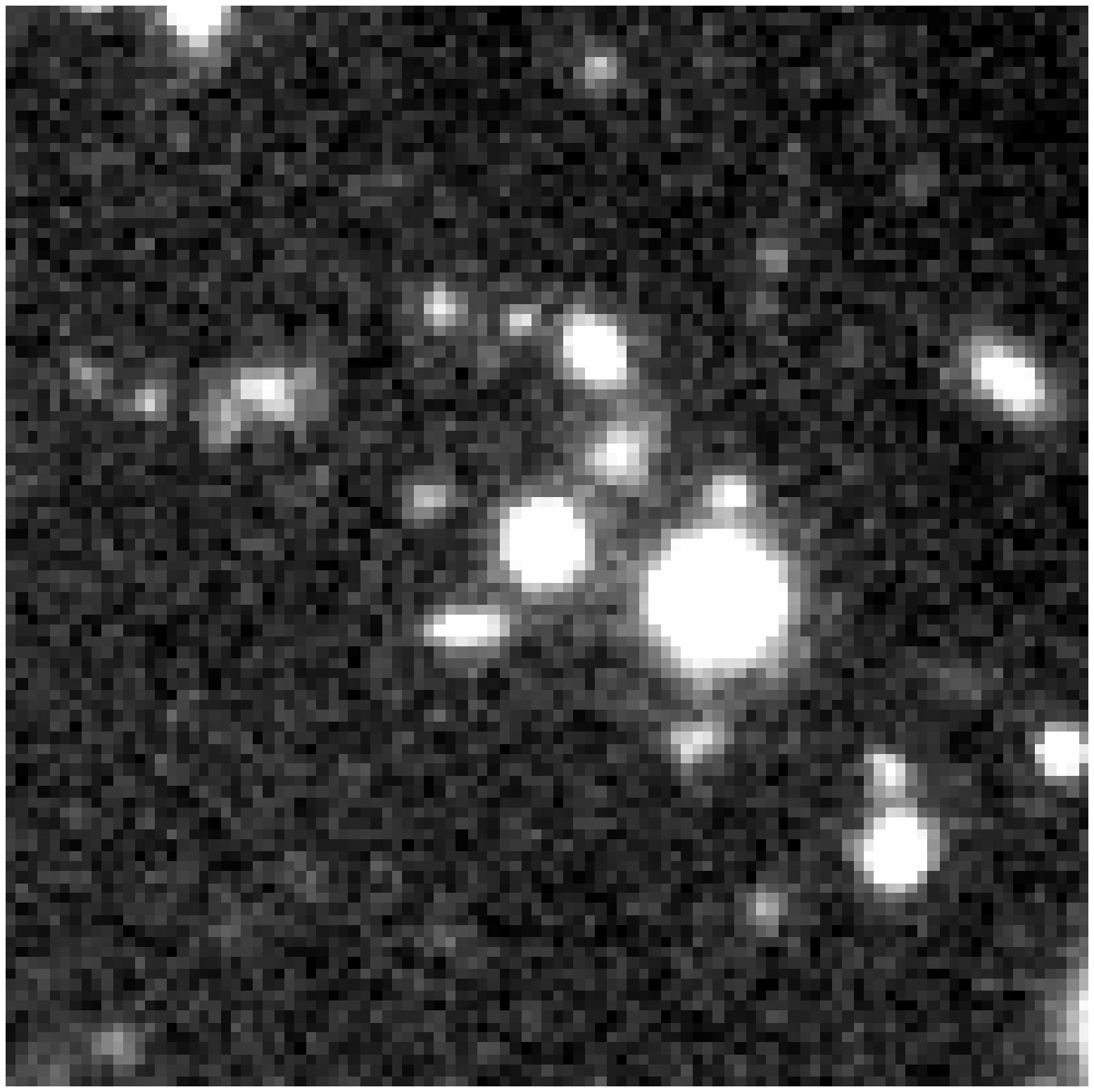,width=3.5cm,clip=t}
\hspace*{0.1cm}
\psfig{figure=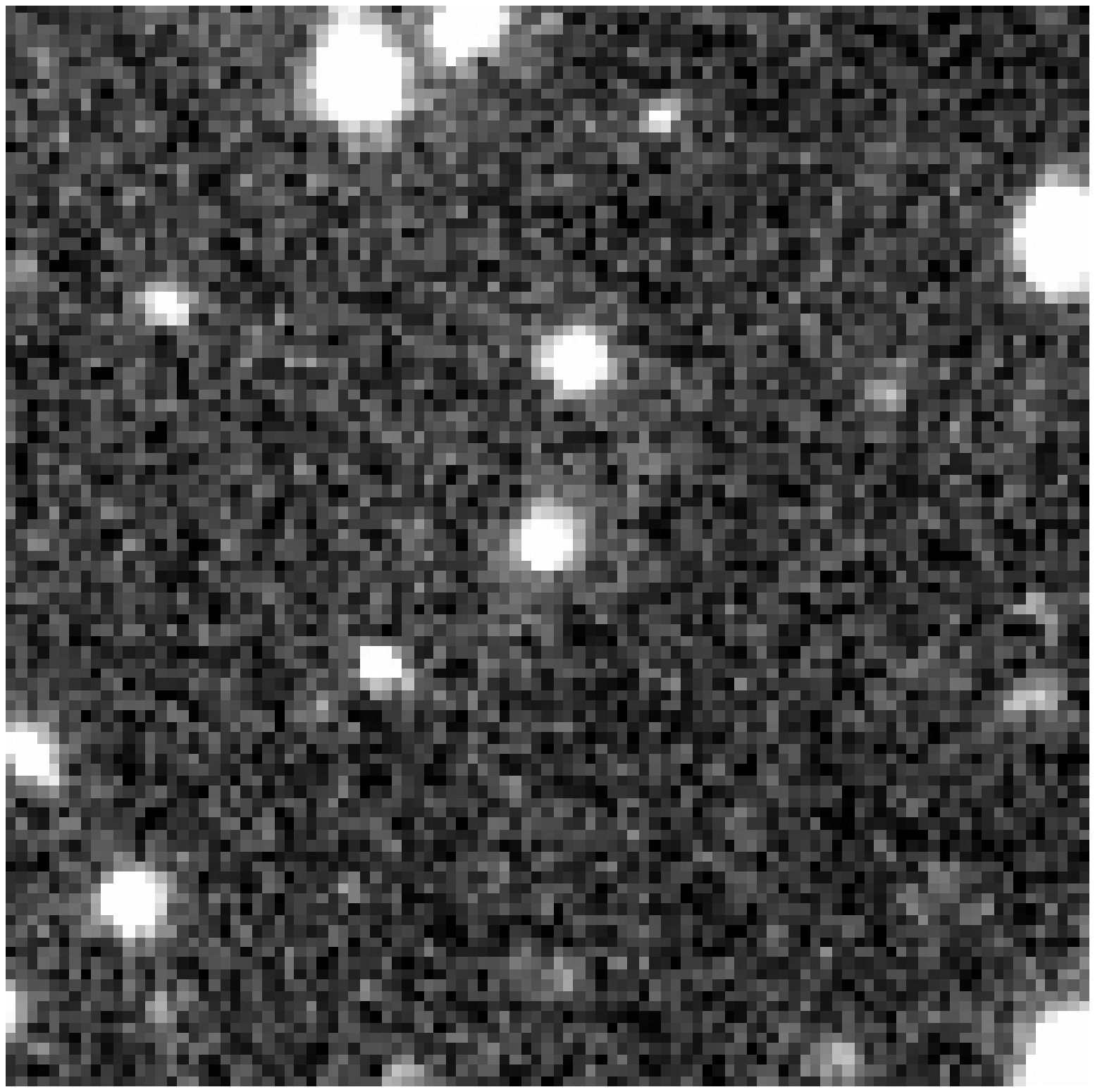,width=3.5cm,clip=t}
\hspace*{0.1cm}
\psfig{figure=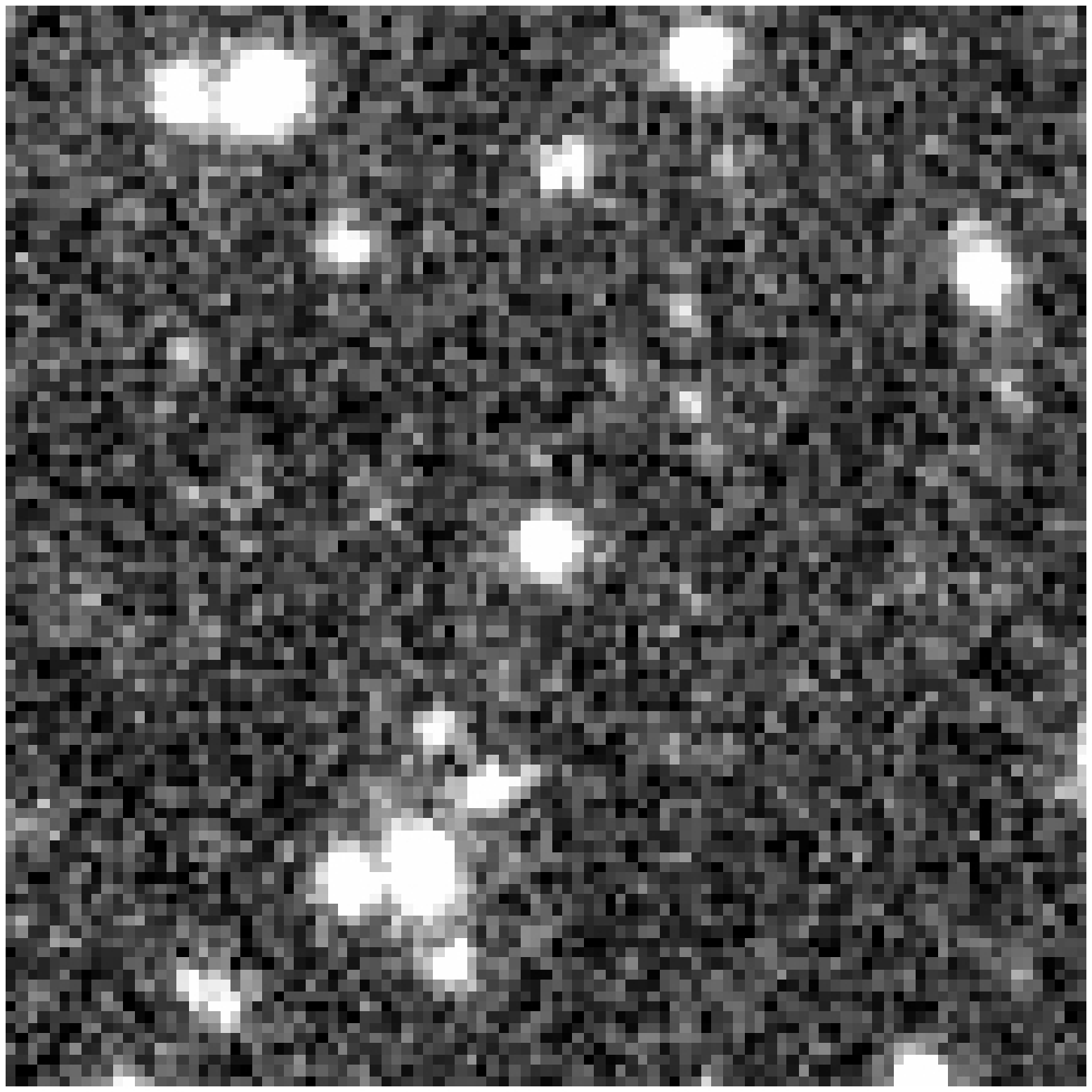,width=3.5cm,clip=t}
}}
\caption{Images of all eight quasars in I. From upper left to lower right: FDF0809, FDF1837, FDF2229, FDF2633, FDF4683, FDF5962, FDF6007, FDF6233. Size of each picture is 20'' $\times$ 20'' respectively, pixelscale is 0.''2/pixel, north is up, east is left.}
\label{pictures}
\end{figure*}

The sample consists of all available eight quasars from the FORS Deep Field (FDF) with redshifts between $z=0.87$ and $z=3.37$.  Except FDF4683, all quasars are radio-quiet (using the criteria by Kellerman et al. (\cite{Ke89}), the radio fluxes (5GHz) are from S. Wagner, private communication). The quasars are identified spectroscopically by Noll et al. (\cite{nollfdf}). Their basic properties can be found in Tab. \ref{qsosample}. Images of all quasars in $I$-band can be found in Fig. \ref{pictures}.
\begin{table*}
\caption{The Quasar-Sample}
\label{qsosample}
\centering
\begin{tabular}{c c c c c c c c}
\hline \hline
FDF-ID & $z$ & RA & DEC & $m_{I}$ & Type \\
\hline
FDF0809 & 0.8650 & 01:05:50.3 & --25:44:48 & 21.4 & RQQ \\
FDF1837 & 2.2540 & 01:05:54.2 & --25:46:41 & 22.9 & RQQ \\
FDF2229 & 2.1560 & 01:05:55.7 & --25:47:22 & 20.8 & RQQ \\
FDF2633 & 3.0780 & 01:05:57.3 & --25:44:57 & 22.8 & RQQ \\
FDF4683 & 3.3650 & 01:06:04.4 & --25:46:51 & 18.6 & RLQ \\
FDF5962 & 1.7480 & 01:06:09.0 & --25:42:56 & 21.9 & RQQ \\
FDF6007 & 2.7515 & 01:06:09.2 & --25:43:26 & 24.1 & RQQ \\
FDF6233 & 2.3215 &01:06:10.0 & --25:43:42 & 23.9 & BAL(RQ) \\
\hline
\end{tabular}
\end{table*}

The field was carefully selected to contain no stars brighter than 18th magnitude but to contain a high-redshift-quasar (FDF4683 at $z=3.37$). The field is about 7' $\times$ 7' in size. Photometry is available for about 9000 objects, spectroscopy for about 350 objects.

The observations were carried out using three different telescopes; the VLT (photometry in UBgRI and spectroscopy), the NTT (photometry in J and Ks) and the HST (F814W photometry). A detailed description of the field selection and the photometry can be found in Heidt et al. (\cite{heifdf}), a description of the spectroscopic observations can be found in Noll et al. (\cite{nollfdf}).

Photometric observations were carried out between August 1999 and August 2000. Integration times were between 6 and 12 hours in the optical bands, about 80 min in the NIR and 40 min with the HST. The image quality of the ground-based data is excellent in the optical with a FWHM between 0."5 (I) and 1" (U) and less good in the NIR ($\sim$1."2). The optical images are extremely deep with 50\% completeness limits between 25.6 (U) and 27.7 (B).

\section{Data Analysis}

\begin{table*}
\caption{Basic PSF properties.}
\label{PSF}
\centering
\begin{tabular}{c c c c c l}
\hline \hline
Filter & pixel-scale [''/pix] & FWHM [pix] & $\beta$ & ellipticity & PA of ellipticity\\
\hline
U & 0.''2 & 4.75$\pm$0.04 & 3.03$\pm$0.02 & 0.04$\pm$0.01 & --8.20$\pm$1.20\\
B & 0.''2 & 2.80$\pm$0.05 & 2.31$\pm$0.03 & 0.04$\pm$0.02 & --339.69$\pm$1.13\\
g & 0.''2 & 4.22$\pm$0.06 & 2.53$\pm$0.04 & 0.04$\pm$0.01 & --195.52$\pm$0.94\\
R & 0.''2 & 3.66$\pm$0.05 & 2.42$\pm$0.01 & 0.06$\pm$0.02 & --24.55$\pm$1.32\\
I & 0.''2 & 2.47$\pm$0.03 & 2.50$\pm$0.04 & 0.03$\pm$0.01 & --35.94$\pm$1.45\\
F814W & 0.''05 & 1.86$\pm$0.02 & 1.75$\pm$0.03 & 0.02$\pm$0.01 & 286.70$\pm$0.86\\
J & 0.''288 & 3.90$\pm$0.05 & 3.59$\pm$0.06 & 0.01$\pm$0.01 & 24.83$\pm$1.13\\
Ks & 0.''288 & 4.18$\pm$0.06 & 3.14$\pm$0.06 & 0.05$\pm$0.01 & 203.93$\pm$0.99\\
\hline
\end{tabular}
\end{table*}

The data were analysed using a 2D-fitting routine based on a Levenberg-Marquardt-loop (see Nilsson et al. \cite{nilsson}). Three different fits were done for each object in each filter: core, core+bulge, core+disk.

For the first fit (core), a point source (core) represented by the Point Spread Function (PSF) was fitted to the quasar to determine the centre coordinates of the object. This fit has three free parameters: the magnitude of the point source and the x- and y-coordinate. Maintaining the determined centre coordinates, a core+galaxy-model was fitted to the object with the galaxy being either a disk with an exponential light profile ($\beta$=1) or a bulge with a de Vaucouleur profile ($\beta$=0.25). Galaxy models are convolved with the PSF. Both the ellipticity and position angle (PA) were set to zero. This fit has three free parameters: the magnitude of the central point source and the magnitude and effective radius of the galaxy. For the object with the lowest redshift (FDF0809 at $z=0.87$) an additional fit with five free parameters (magnitude of central point source, magnitude, effective radius, ellipticity and position angle (PA) of the galaxy) was done because the host galaxy of this QSO was clearly resolved on the original images already. Since bluewards of the Lyman break, almost no light from the host galaxy is to be expected, the core+galaxy-fit should not yield a detection of the host galaxy for FDF2633, FDF4683 and FDF6007 in U and FDF4683 in U and B. Nevertheless, core+galaxy fits were also carried out in these cases as a critical test of the host galaxy detection procedure.

The construction of the PSF is a very critical point, as for such high redshifts and only marginally extended host galaxies the PSF has to be known very precise. This was a problem here, as the FDF by selection contains no bright stars and there is a risk that the wings of the PSF are not properly determined.

For example, the apparent magnitudes of the useful PSF-stars in the $I$-band lie between 20 and 22. For three of the eight quasars the apparent magnitudes are in the same range, two have apparent magnitudes of about 23, two have apparent magnitudes of about 24. The PSF works well for objects that are slightly fainter than the PSF stars. PSF-stars that are as faint as the faintest objects were not used as their signal-to-noise is too low. Likewise, the three brightest PSF-stars (19--20 mag) could not be used due to blending with other objects. This was especially important for the object FDF4683 at $z\approx3.4$ which is the second brightest object in the field (the brightest object is an extended elliptical galaxy). With an apparent magnitude of 18.6, FDF4683 is at least 3 magnitudes brighter than all available PSF-stars.

Applying these selection criteria and depending on the filter, $\sim10$--$20$ stars were used to build a PSF. A preliminary PSF was built out of all PSF-stars and improved iteratively by fitting the PSF to the all available stars and rejecting apparently bad PSF-stars, meaning PSF stars that showed ``double-source structure'' or very high ellipticity. No systematic variations of the PSF across the FOV was found, thus one PSF per filter was used for the fits. We checked the variations over across the field by fitting PSFs build from nearby stars and ``whole-field-PSFs'' to each quasar. The results agreed well within the error bars. Thus we decided to use the PSF build from stars all over the field due to the higher signal to noise ratio.

Additionally, fitting the PSF to the available stars gives a reliability check for the fitting method. I.e. if this kind of fits yields host galaxy detections, the method is hardly reliable. In that case, spurious host galaxy detections are expected. However, fitting the PSF to stars, we were not able to detect spurious host galaxies for FORS data. As for the quasars, we fitted a core and a core+galaxy model to the field stars. The core fits yielded $\chi^{2}\approx1$ and core+galaxy fit caused the effective radius to converge to zero which is a clear sign that no host galaxy can be resolved.

For the HST-data it was much more difficult to build a PSF, as its PSF shows strong variations across the field. After various attempts, we finally used a PSF kindly provided by Maurilio Pannella (MPE Garching), see Pannella et al. (\cite{Pa06}). This PSF was built by stacking all the PSF-stars in the field. This PSF does not describe the PSF-variations across the field, however it is the only method that yields reliable results. Testing other PSFs, spurious host galaxy detection in PSF stars were found, thus we rejected the affected PSFs as not reliable.

The basic properties of our PSFs can be found in Tab. \ref{PSF}. The values were determined by fitting an ellipse with 4 free parameters to the PSF. These values, including the error bars are used for further error simulations.

We performed error simulations for objects where the $\chi^{2}$ for one of the core+galaxy-fits was significantly better than for the core-fit.

For the error simulations, the object is described using the results of the fits and convolved with a simulated PSF with noise added. The model PSF is built using various parameters of the used PSF including the errors of these parameters, derived by fitting a Moffat function to the PSF. Afterwards, the program fits the simulated objects. Thus, this method includes detector errors, shot noise, errors of the fitting procedure and errors of the PSF. As this procedure involves fitting only synthetic galaxies with the best galaxy model, this does not include uncertainty from the mismatch of the model to the real galaxy shape.

Approximately $20$--$50$ simulations are done for each object. The criteria for the object being resolved or not is: $5\times\sigma_{r_{e,simulated}}<r_{e,fits}$, with $\sigma_{r_{e,simulated}}$ being the standard deviation of the effective radius from the simulations and $r_{e,fits}$ being the measured effective radius. This has to be the case at least in one filter for a particular object. Additionally we checked the resulting fitting images from the error simulations, in case those are reliable, the fitting results from the error simulations should look similar to normal fitting results. Residuals from the error simulations were found to be similar to normal residuals. Thus we rely on the error simulations to reproduce the data well.

Additionally, we consider the absolute values of the effective radii in pixels. Resolved host galaxies with effective radii on sub-pixel scales are highly suspicious. We discuss this point.

\section{Results}

In this Section, we will present the results. After discussing which galaxies were resolved, we will estimate the galaxy types and the approximate ages based on the host galaxy colours and will derive the star formation rates of the resolved host galaxies. Finally we will derive the mass of the central black hole of the QSOs.

\begin{figure*}[t]
\centering
\includegraphics[width=15cm]{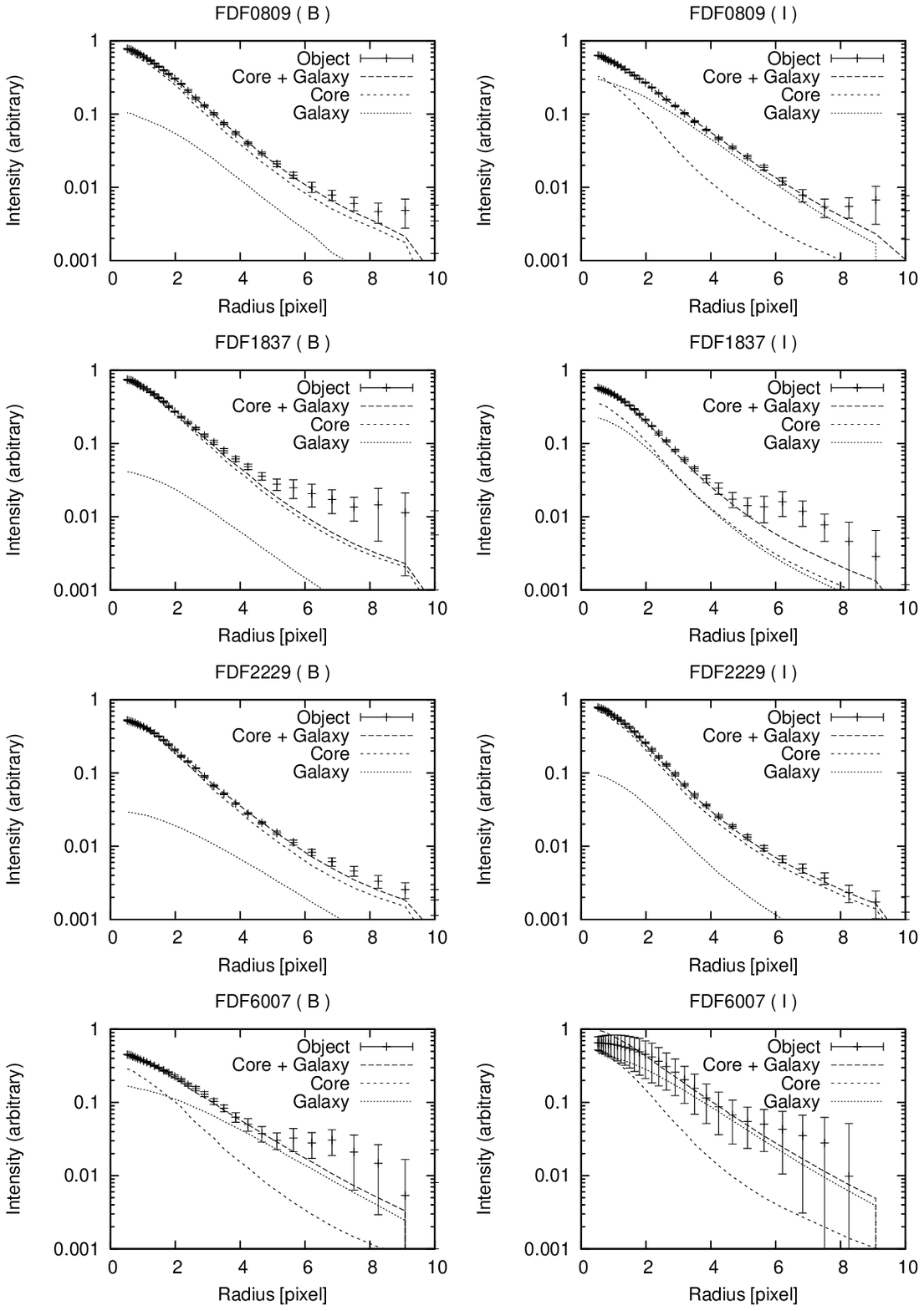}
\caption{Fitting results of the resolved and marginally resolved objects, see text for discussion. From top to bottom: FDF0809 (ellipticity and PA fitted), FDF1837, FDF2229, FDF6007. Left panel: B band, right panel: I band. All intensities are plotted logarithmically and in arbitrary units. Radius in pixels.}
\label{resolvedobjects}
\end{figure*}

\begin{figure*}[t]
\centering
\includegraphics[width=15cm]{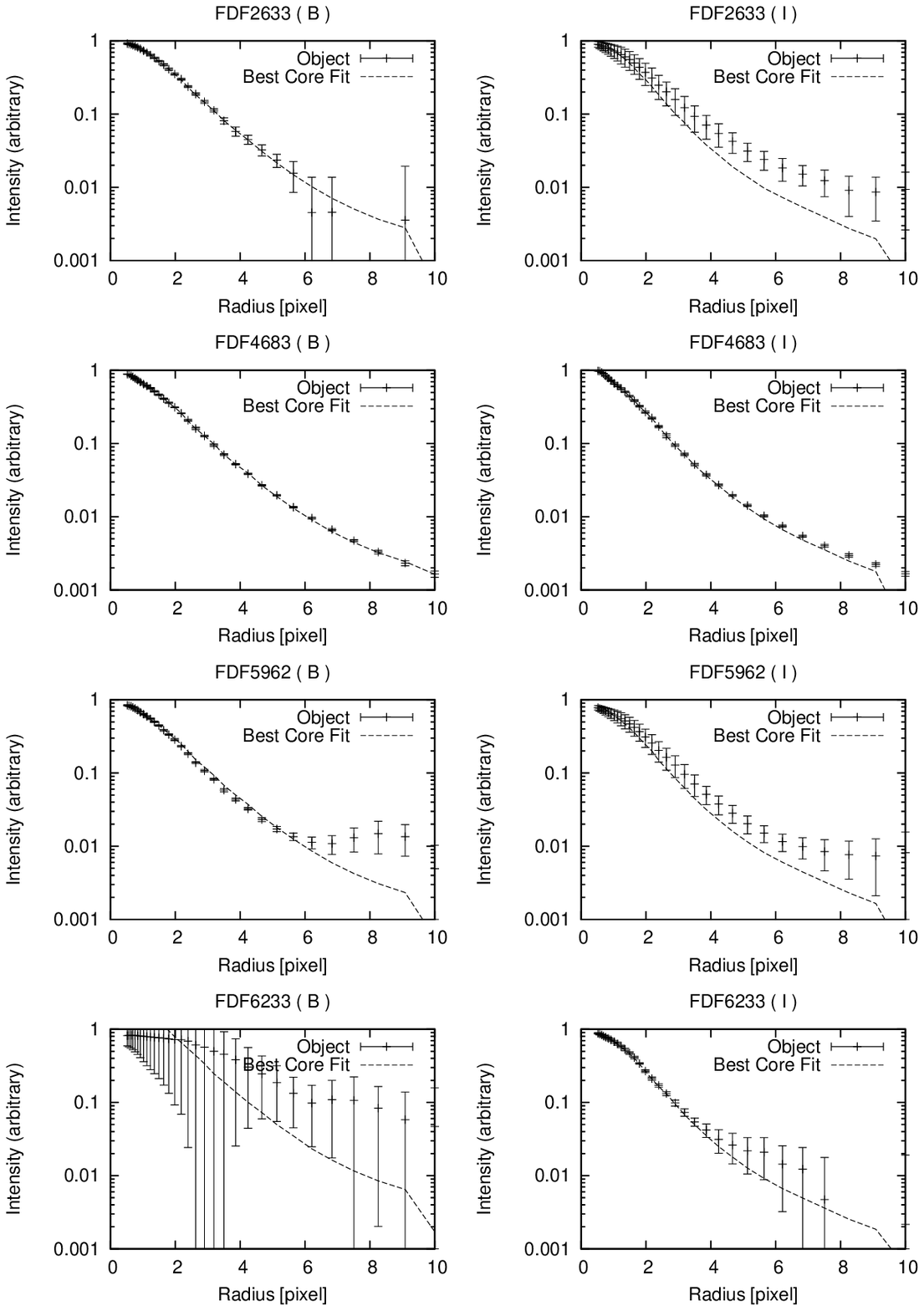}
\caption{Fitting results of unresolved objects. From top to bottom: FDF2633, FDF4683, FDF5962, FDF6233. Left panel: B band, right panel: I band. All intensities are plotted logarithmically and in arbitrary units. Radius in pixels.}
\label{unresolvedobjects}
\end{figure*}

The host galaxy was resolved in two out of eight quasars (FDF0809 and FDF6007). FDF0809 was resolved in all filters. FDF6007 was resolved in BVRI, no fits were possible in UJKs an F814W due to the faintness of the object. FDF1837 and FDF2229 were possibly resolved, we will discuss this in the following paragraphs. The rest of the objects are clearly unresolved.

The results of the best fits to each object in the I-band are shown in Fig. \ref{resolvedobjects} (all resolved and marginally resolved objects) and Fig. \ref{unresolvedobjects} (all unresolved objects). In Table \ref{M-result}, the absolute magnitudes for the resolved host galaxies are given, values for FDF0809 are from the fit with variable ellipticity and PA. K-corrections for FDF0809 are from Bicker et al. (\cite{Bi04}). Note that we did not apply k-corrections for the objects FDF1837, FDF2229 and FDF6007, as for this high redshift that would mean interpolating from rest-frame ultraviolet to rest-frame optical and thus from young to old stellar populations. Thus the magnitudes are not given in the filters specified but in the concerning filters transferred to the redshift of the object. Cosmology is $H_{0}=70\textrm{km/s/Mpc}, \Omega_{\Lambda}=0.7, \Omega_{m}=0.3$, unless for evolution-corrected magnitudes of host galaxies, where cosmology is: $H_{0}=65\textrm{km/s/Mpc}, \Omega_{0}=0.1$. The tables with all the fitting results are presented in Appendix A. We now comment on each object separately, host galaxy radii will be discussed in greater detail in the following section.

\begin{table*}
\caption{Absolute magnitudes for the resolved host galaxies. We marked the magnitudes for objects for which the radii indicate that the galaxy is only marginally resolved. For a full result table, see Appendix A. For further information, see discussion in the text.}
\label{M-result}
\centering
\begin{tabular}{l c c c c c c c c c c}
\hline \hline
object & $M_{U}$ & $M_{B}$/$m_{B}$ & $M_{g}$ & $M_{R}$ & $M_{I}$/$m_{I}$ & $M_{F814W}$ & $M_{J}$ & $M_{Ks}$ & $r_{B}$[``] & $r_{I}$[``]\\
\hline
FDF0809 [disk] & --21.02 & --20.02/24.62 & --21.32 & --22.02 & --22.22/21.79 & --22.26 & --22.62 & --23.62 & 0.24 & 0.32 \\
FDF0809 [bulge] & --24.12 & --22.52/24.63 & --22.52 & --23.12 & --23.12/21.43 & --23.34 & --22.62 & --23.32 & 0.21 & 0.26 \\
\hline
FDF1837 [bulge] & --21.99 & --19.95/26.32 & --19.69 & --21.09$<$ & --22.23/24.04$<$ & -- & -- & -- & 0.30 & 0.11 \\
FDF1837 [disk] & --21.79 & --19.78/26.49 & --21.36 & --21.66$<$ & --22.48/23.79$<$ & -- & -- & -- & 0.26 & 0.12 \\
FDF2229 [bulge] & --23.40 & --22.35/23.81 & --22.62 & --23.14$<$ & --23.09/23.07$<$ & -- & -- & -- & 0.31 & 0.13 \\
FDF2229 [disk] & --23.63 & --21.88/24.28 & --22.06 & --23.11 & --23.37/22.79$<$ & -- & -- & -- & 0.38 & 0.13 \\
FDF6007 [bulge] & -- & --21.60/25.20 & --20.81 & --22.15$<$ & --22.31/24.49 & -- & -- & -- & 0.50 & 0.67 \\
FDF6007 [disk] & -- & --21.10/25.70 & --20.78 & --21.74 & --21.76/25.04 & -- & -- & -- & 0.44 & 0.54 \\ 
\hline
\end{tabular}
\end{table*}
 
FDF0809 ($z=0.8650$): This is the object with the lowest redshift, the host galaxy was resolved in all filters. Error simulations yielded resolution with 5$\sigma$ significance in six out of eight filters and resolution with 3$\sigma$ significance in the two other filters. Thus the host galaxy is clearly resolved and has a disk type morphology. The radii found are in the range $0.''2$--$0.''5$, corresponding to $1.5$--$4$kpc. The radii are significantly smaller than the mean seeing, still, they are all larger than one pixel, thus we consider the galaxy as resolved.

FDF1837 ($z=2.2540$): A host was formally detected in the filters UBgRI. According to the error simulations, the object is resolved in the filters B (5$\sigma$) and I (3$\sigma$). With effective radii ranging from $0.''11$--$0.''30$ and a pixel scale of the chip of $0.''2$ we would like to point out that at least for the filters R and I the galaxy doesn't seem resolved. Taking the error bars for $r_{e}$ into account we are on the limit of rejecting the host galaxy detection due to sub-pixel host galaxy radii. In this situation, even a 5$\sigma$-significance is not a strong argument for the host galaxy being resolved. Thus, the magnitudes are considered to be upper limits. The object has a close companion in the north, it is not clear if it is associated to tidal tails, a neighbouring interacting galaxy or a interloper. The object was masked for fitting but is clearly visible in the plots in Fig. \ref{resolvedobjects}. This may lead to problems in the error simulations as the neighbouring object is not modeled, thus heavily underestimating errors caused by the influence of this object. Additionally, we would like to point out that the filters with host galaxy detections on scales $r_{e}>1$pix (UBg) are affected by broad emission lines. In case of extended quasar emission, we might not have detected the host galaxy, but other kinds of extended emission. For example, Davies et al. (\cite{Da08}) detected extended line emission in high redshift galaxies on scales of $\sim4$kpc, this would be well consistent with the ``host galaxy detections'' in FDF1837, being on scales of $\sim2$--$4$kpc. All in all the detection is highly questionable. Still, we perform all upcoming tests on this object, but all the arguments mentioned above should be kept in mind.

FDF2229 ($z=2.1560$): A host was formally detected in the filters UBgRI. The error simulations yielded a $3\sigma$ significance in the U and B-band and a $5\sigma$-significance in the g-band. As for FDF1837, red bands here show very small effective radii, we therefore follow the arguments used for FDF1837. We consider the host galaxy as possibly resolved. The magnitudes are considered to be upper limits. As for FDF1837, broad emission lines are present in all filters with host galaxy detections on scales $r_{e}>1pix$ (UBg).

FDF2633 ($z=3.0780$): This is the object with the second highest redshift, it has a close neighbour to the north-west. As the close neighbour is a large elliptical galaxy, it is possible that the outer parts of this galaxy influence the fits. In order to improve the quality of the fits we iteratively removed the contribution of this neighboring galaxy as described in Heidt et al. (\cite{He99}). Nevertheless we were not able to resolve the host galaxy.

FDF4683 ($z=3.3650$): This is the second brightest object in the FDF and the quasar with the highest redshift. Due to the difficulties in extracting a suitable PSF as described above, the host galaxy could not be resolved.

FDF5962 ($z=1.7480$): As in the case of FDF2633 we tried to remove the contribution of the neighboring objects iteratively, but could not resolve the host galaxy.

FDF6007 ($z=2.7515$): This is the object with the highest redshift where we could resolve the host galaxy. The object was formally resolved in the filters BgRI. No fits were possible in the filters UJKs due to the object being too faint. According to the error simulations, the object was resolved with $5\sigma$-significance in R and I and with $3\sigma$-significance in B. However, $Ly\alpha$ lies in the B band, thus the same argument holds as for FDF1837 and FDF2229. With an effective radius of $3.9/3.5$kpc for the different fits respectively in B this is well consistent with extended gas emission detected by Davies et al. (\cite{Da08}). The other bands show partially very small radii or implausibly small error bars, such as $\pm0.''02$ in R band. Additionally, this object lies in between the classification for TypeI an TypeII quasars, thus the core component might be obscured, making the measured magnitudes lower limits. All in all, the galaxy is considered to be marginally resolved with somewhat uncertain magnitudes due to possible obscuration and small radii.

FDF6233 ($z=2.3215$): This object is a broad absorption line quasar, it could not be resolved.

\subsection{Host galaxy radii}

Only few studies of high redshift quasar host galaxies have published radii. Kukula et al. (\cite{Ku01}), who observed in J and Ks, found a mean of 11.1$\pm$5.7 kpc for their $z\approx1$ sample. For their $z\approx1.5$--$2$ sample they found a value of 9.78$\pm$6.58 kpc for objects where it could be determined and upper limits ($<$10 kpc) for the remaining sources. Kotilainen et al. (\cite{Ko07}) found values of 6.7$\pm$1.7 kpc for their $z\approx1$--$2$ sample (observed in K and H). Falomo et al. (\cite{Fa08}) found values between 2.6 kpc and 11.3 kpc for a sample of three high redshift quasars ($z\approx2$--$3$) observed in Ks.

In this study we find the following values for our host galaxies: FDF0809 at $z=0.9$ which was clearly resolved has radii between $1.5$--$3.9$kpc in different bands. FDF6007 at $z=2.8$ which was clearly resolved shows radii in the range of $1$--$5.3$kpc. For the possibly resolved objects the radii are the most critical point, thus they will not be discussed here.

Our values are in general lower than the ones found by other authors. We would like to point out some effects that might cause these differences. For FDF6007, we measured the effective radii in rest-frame UV, whereas all other values were measured in rest-frame optical to rest-frame NIR. Thus the observed trend might be expected if the galaxies are bulge-dominated. Moreover the preferred morphology in the cited studies is a de Vaucoleur profile. FDF0809 was identified as a late type galaxy and late type host galaxies in our sample might be expected as the quasars we studied are relatively faint, whereas most other studies concentrate on bright targets. For low redshift host galaxies clear correlations between quasars and their host galaxies were found, with fainter quasars residing in relatively faint late type galaxies and brighter quasars residing in relatively bright elliptical galaxies. As all our quasars for which we were able to resolve the host galaxy are on the faint end of the quasar population, we expect faint, disk dominated galaxies. The two studies that concentrate on high redshift, low-luminosity quasars (Jahnke at al. \cite{Ja04}; Ridgway et al. \cite{Ri01}) did not specify radii due to a different fitting procedure. Additionally, if we only consider the radii larger than one pixel, i.e. the cases in which we regard the galaxy to be clearly and not only possibly resolved, this corresponds to $r_{e}\approx2$--$3$kpc which is consistent with smaller galaxies and even lies in the lower range of galaxies resolved by other authors. For FDF0809, the $I$-band corresponds to rest-frame optical, we find radii of $2.0/2.5$kpc depending on morphology, this lies in the lower range of values found by other authors but is sensible considering that the galaxy was identified as a disk-dominated galaxy.

Additionally, Hutchings (\cite{Hu06}) has pointed out that high redshift quasar host galaxies are found to be irregular and highly disturbed. He also argued that the fact that high redshift quasar host galaxies were mostly found to be elliptical might be caused by the fact that only the bulges were detected. This could also explain relatively small radii, simple meaning that we only resolved the centre bulge and not the probable more extended structures.

Considering all these arguments, the radii we measured seem sensible.

\subsection{Simulation of spectra}

\begin{figure*}
   \centering
   \includegraphics[width=5cm,angle=270]{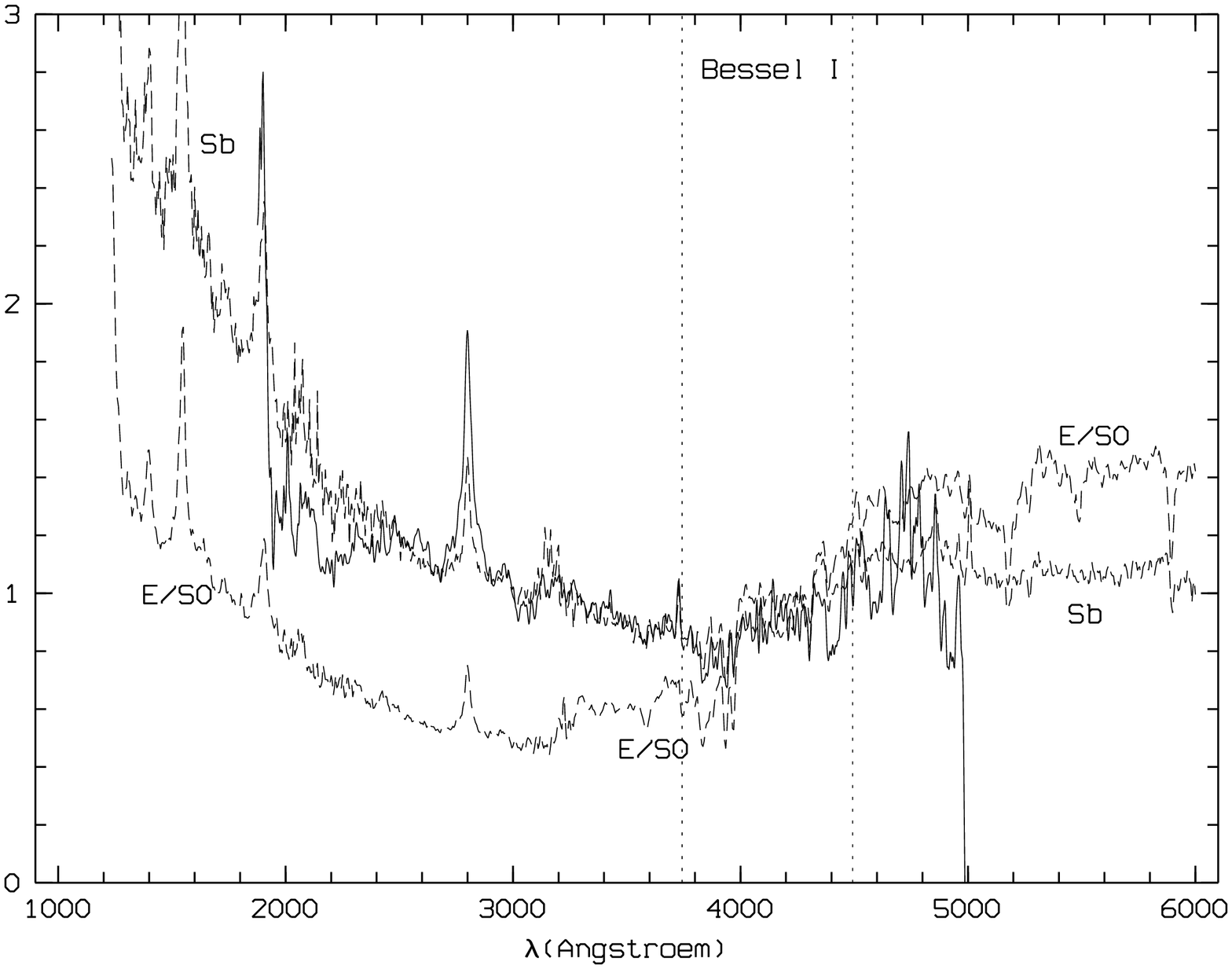}
	\hspace{1cm}
   \includegraphics[width=5cm,angle=270]{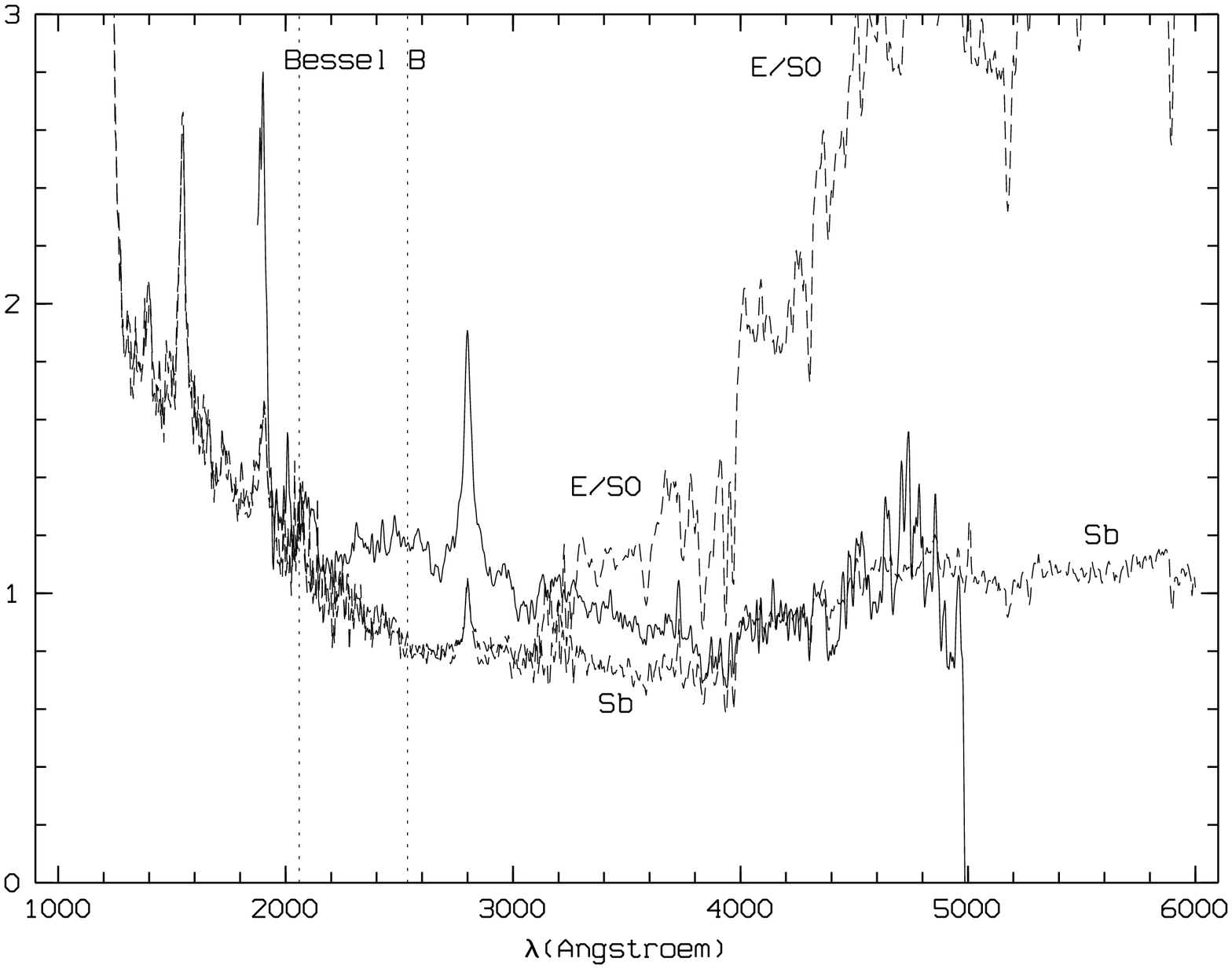}
      \caption{Simulation of the spectrum of FDF0809 in the I-band (left) and B-band (right). The solid line shows the measured spectrum in the restframe, the dashed lines show the simulated spectra (QSO+galaxy spectra added up) marked with the model type of the galaxy template. The wavelength range used for scaling is marked with dotted lines and the name of the corresponding filter. The Sb-model is unambiguously preferred in both cases.}
         \label{spectrsimu}
   \end{figure*}

In order to investigate the dominant stellar populations we analysed the spectrum of FDF0809 from the FDF data (see Noll et al. \cite{nollfdf}) as follows:

The expected flux ratio of core to galaxy from the fits was calculated for a virtual 1" slit (corresponding to FORS spectroscopy performed for our object) using the model images of the core and the galaxy in the B and I bands. We used the QSO-template spectrum by Walsh, based on Francis et al. (\cite{Fr91}), and an elliptical galaxy and a spiral Sb galaxy template spectrum, both by Kinney et al. (\cite{Ki96}), all of which were folded with the transmission curves of the used filter, and then finally transformed to the restframe of the object. The spectrum was then scaled to flux = 1.0 in this virtual filter. The calibrated spectra were weighted with the flux of the central point source and the host galaxy, respectively. Finally, the two weighted spectra were added up and compared to the spectrum of the object. In both cases, a stellar population of a late type is clearly favoured over a stellar population of an early type galaxy (see Fig. \ref{spectrsimu}).

Letawe et al. (\cite{Le06}) and Canalizo \& Stockton (\cite{CS01}) studied low-redshift, low luminosity quasars and found late type stellar populations with recently induced starburst. Thus our findings are well consistent with those of other authors.

We also tried to simulate the spectra of the other resolved QSO hosts. Due to the fact that these other objects lie at higher redshifts, for these objects (FDF1837, FDF2229 and FDF6007) we sample a redshift range where the different galaxy spectra cannot be distinguished sufficiently from each other. Thus our attempt was not successful.

\subsection{Investigation of galaxy type and age using colours}

To investigate the galaxy type and the redshift of the formation of the galaxy, we used galaxy models based on Bruzual \& Charlot (\cite{BG93}) with solar metalicity and a formation of the host galaxies at $z=3$ and $z=5$. The apparent colours of different model galaxies as a function of redshift are compared to the colours of our resolved host galaxies.

\begin{figure*}
   \centering
   \includegraphics[width=5cm,angle=270]{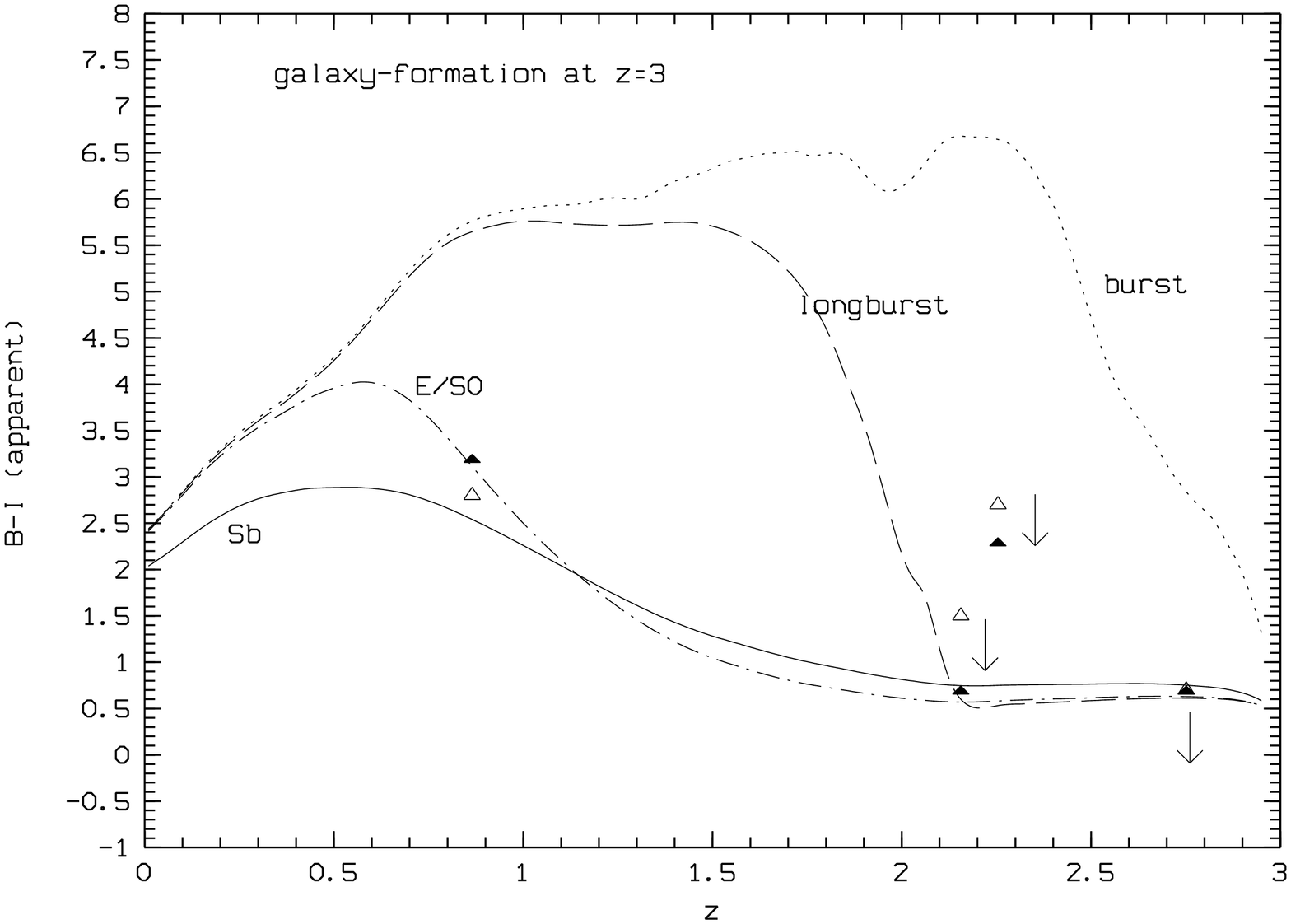}
	\hspace{1cm}
   \includegraphics[width=5cm,angle=270]{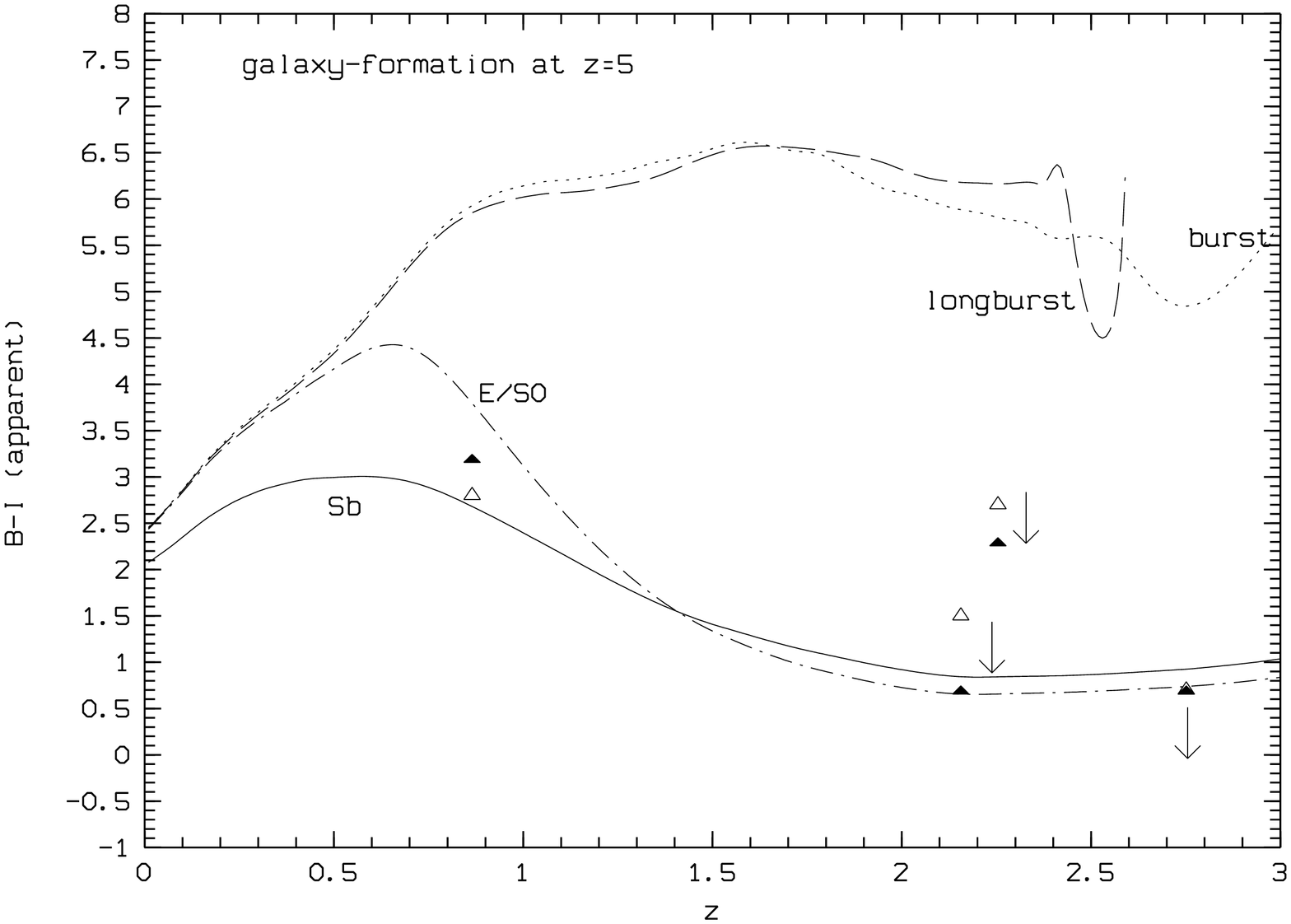}
      \caption{Colours (B--I) of model galaxies as a function of redshift compared to the colours of our resolved galaxies. The redshift of the galaxy formation is $z=3$ (left) and $z=5$ (right) respectively. Filled triangles indicate the results of our fits with an elliptical galaxy and open triangles with a disk galaxy. The colours of burst galaxies are shown as dotted, the ones of long burst galaxies as long-dashed, the ones of E/S0 galaxies are shown as dash-dotted lines and the ones of Sb galaxies as solid lines, respectively. If a filter is influenced by forbidden or semi-forbidden lines, originating in the BLR/NLR the magnitude of the host galaxy in this filter can be overestimated, as the flux from these lines may be detected as host galaxy flux. This is indicated by arrows. Errors lie in the order of $\sim1mag$}
	\label{colplot}
   \end{figure*}

We use following galaxy models: A model where all stars form in an instantaneous burst, afterwards the stellar populations evolve passively (''burst''); a model where all stars formed in a starburst lasting 1 Gyr, afterwards the stellar populations evolve passively (''longburst''); an E/S0 type stellar population, described by a star formation rate is proportional to exp(-t/1Gyr) (''E/S0) and a Sb type stellar population. The Initial Mass Function (IMF) used is Salpeter IMF for burst, longburst and E/S0 and a Scalo IMF for Sb. This was done for B--V, B--I, V--R, R--I and I--J. If a filter is influenced by forbidden or semi-forbidden lines, originating in the broad or narrow line region the magnitude of the host galaxy in this filter can be overestimated, as the flux from these lines may be detected as host galaxy flux. Though, this is a problem only if the emitting gas is extended significantly. The results for the B-I colour (which gives the largest wavelength baseline) are shown in Fig. \ref{colplot}.

Note that the naming is somewhat confusing, i.e. both burst and longburst form stars only for a very short time period, afterwards there is no ongoing star formation.

For the object FDF0809 ($z=0.865$) the colours are compatible with both a E/S0 or Sb-galaxy and both a formation of the galaxy at $z=3$ or $z=5$. Due to minimal deviations it is not possible to specify the age or galaxy type further. Starburst galaxies can be excluded unambiguously. For the other object it is more difficult to specify the galaxy type, as the colours for models Sb, ES0 and longburst with a galaxy formation at $z=3$ show merely no deviations. Thus for FDF6007 the colours are compatible with galaxy formation at $z=3$ and the models E/S0, Sb and long burst or a galaxy formation at $z=5$ and models E/S0 and Sb.

For FDF1837 and FDF2229 the magnitudes are merely upper limits. Considering this fact, the colours are compatible with galaxy formation at $z=3$ and the models E/S0, Sb and long burst or a galaxy formation at $z=5$ and models E/S0 and Sb. Still, the errorbars might be very high.

\subsection{Star Formation Rate}

To estimate the star formation rate (SFR) of the resolved host galaxies we used the method advocated by Kennicutt (\cite{Ke98}). Using the flux in the UV-continuum between 1250 and 2500 \AA{}, the SFR can be calculated as:
\begin{displaymath}
 SFR(M_{\odot}/yr)=1.8\times10^{-27} \left\lbrace \frac{d_{l}^{2}\times10^{-0.4(m_{AB}+48.6)}}{1+z} \right\rbrace 
\end{displaymath}
with $d_{l}$ the luminosity distance and $m_{AB}$ the apparent magnitude in the AB-system. The two objects FDF1837 and FDF2229 suffer from strong semi-forbidden lines originating from the AGN in the relevant filter. These lines originate from the narrow or broad-line regions but may be detected as host galaxy flux in case these regions are extended significantly. For those objects, the SFR using the fitted magnitudes gives an upper limit. To derive a lower limit for the SFR we subtracted the line-flux (derived from the spectra of the particular object) from the flux of the galaxy. Still, one has to take into account that for FDF1837 and FDF2229 all extended emission might be caused by gas and not the host galaxy.

Note that SFRs may be underestimated using this method as the possible presence of dust has not been taken into account.

Unfortunately, this method does not yield convincing results if the SFR changes on time scales shorter than $\sim10^{8}$yr (actually, this may well be the case for very young or actively star-forming galaxies!). Therefore, as a consistency check, the SFR was determined not only for the resolved galaxies but additionally for the model galaxies (based on Bruzual \& Charlot \cite{BG93}) used in the previous subsection, to check the reliability of the results.

This was done as follows: we used the apparent magnitudes from model galaxies at the same redshift and in the same band as the corresponding host galaxies, thereby simulating an observations of a galaxy with the assumed evolution. Then we used the same formula to calculate the SFR as for the host galaxies.

 The results for both the host galaxies and the model galaxies are summarised in Table \ref{SFR}.
\begin{table}
\caption{Star formation rate of the resolved host galaxies and model galaxies in $M_{\odot}/yr$.}
\label{SFR}
\centering
\begin{tabular}{c c c c c}
\hline \hline
object(filter)$\rightarrow$ & 0809(B) & 1837(R) & 2229(R) & 6007(I) \\
\hline
object(bulge fit) & 1.8 & $<$0.03-4.7 & $<$15.8-33.0 & 6.9 \\
object(disk fit) & 1.8 & $<$0.07-8.2 & $<$14.9-30.1 & 4.4 \\
\hline
burst($z=3$) & 0.4 & 0.07 & 0.06 & 15.2 \\
burst($z=5$) & 0.4 & 0.04 & 0.04 & 0.07 \\
long burst($z=3$) & 0.5 & 391.2 & 300.9 & 288.8 \\
long burst($z=5$) & 0.3 & 0.05 & 0.05 & - \\
E/S0($z=3$) & 6.5 & 170.8 & 157.9 & 240.3 \\
E/S0($z=5$) & 2.6 & 68.0 & 57.3 & 87.2 \\
Sb($z=3$) & 5.4 & 8.2 & 8.3 & 6.6 \\
Sb($z=5$) & 4.9 & 6.8 & 6.9 & 6.0 \\
\hline
\end{tabular}
\end{table}

Indeed, the formula to derive the SFR does not work properly for the model galaxies with a rapidly changing SFR. In these cases the SFR is heavily overestimated. It seems to work well for moderate star formation. The values for FDF0809 seem reasonable, especially since both our fits and simulated spectra indicate a disk-type stellar population. On the other hand, rapidly changing SFRs seem reasonable for the high redshift hosts, for which one would expect very young stellar populations, maybe still in the state of heavy star formation.

Jahnke et al. (\cite{Ja04}) used the same method to determine SFRs of their high redshift ($z\approx1.8$--$2.6$) quasar host galaxies sample and found values of about $2$--$30 M_{\odot}/yr$, with a mean of $\sim12M_{\odot}/yr$ (Knud Jahnke, private communication). This is comparable to the mean SFR of Lyman break galaxies ($\sim7M_{\odot}/yr$) at $z=2.5$ (Jahnke et al. \cite{Ja04}). This also agrees well with our findings.

\subsection{Mass of the central black hole}

\begin{table*}
\caption{Mass of the central black hole in $log(M_{BH}/M_{\odot})$ derived from the CIV line width.}
\centering
\begin{tabular}[h]{ccccccccc}
\hline \hline
Object-ID & FDF1837 & FDF2229 & FDF2633 & FDF4683 & FDF5962 & FDF6007\\
\hline 
$log(M_{BH}/M_{\odot})$ & 8.32 & 8.79 & 7.42 & 8.56 & 8.94 & $>$7.04 \\
\hline
\end{tabular}
\label{MBH}
\end{table*}

In the last years, several methods have been developed to derive the mass of the central black hole of AGN empirically. Those are e.g. the correlations between black hole mass and properties of the host galaxies summarised by Novak et al. (\cite{No06}) and the use of the C IV line width discussed in Vestergaard \& Peterson (\cite{VP06}). In the following we will use the latter method to estimate the mass of the central black hole.

The relation uses the C IV emission line width and the flux at rest frame 1350\AA{}. The relationship is empirical and is based on a sample of 32 AGN for which reverberation-based mass estimates are available (Peterson et al. \cite{Pe04}). The mass of the central black hole is determined using:

\begin{displaymath}
log(M_{BH}/M_{\odot})=
\end{displaymath}
\begin{displaymath}
log \left\lbrace  \left( \frac{FWHM(CIV)}{1000km/s} \right) ^{2} \times \left( \frac{\lambda*L_{\lambda}(1350\AA{})}{10^{44}erg/s} \right) ^{0.53} \right\rbrace +\left( 6.66\pm0.01 \right) 
\end{displaymath}

We estimated the mass of the central black hole for all but two quasars: FDF0809, where the C IV emission line was not in the wavelength range covered by the spectrum and FDF6233 which is a BAL quasar and thus does not show strong broad emission lines. The results are displayed in Tab. \ref{MBH} and are between $\sim10^{7}$--$10^{9}M_{\odot}$. Due to the fact that FDF6007 is a in between the classification of a type1 and a type2 quasars, the line width and corresponding flux might be affected by obscuration, thus the determined mass for FDF6007 is a lower limit. The systematic uncertainty for this method is approximately a factor of 4.

Vestergaard et al. (\cite{Ve08}) published black hole masses derived using the same method for 14 584 quasars ($z=0.2$--$5$) from SDSS DR3 (Sloan Digital Sky Survey Data Release 3). The black hole masses from our sample are in the lower range of the evolution seen in the SDSS quasar sample. Though, it should be noted that the SDSS is far less deep than the FDF and thus faint high-redshift quasars comparable to those from our sample are not detected. As the fainter quasars are expected to have lower black hole masses, the fact that our sources show somewhat small black hole masses is sensible.

\subsection{Black hole mass for FDF0809: $M_{BH} - L_{galaxy}$ correlations}

As we cannot derive the mass of the central black hole for FDF0809 using reverberation based methods, we used $M_{BH}-L_{galaxy}$ correlations summarised by Novak et al (\cite{No06}). These methods use correlations between different properties of the galaxy and the central black hole mass, derived from local galaxies were the black hole mass can be measured directly. We use correlations between the black hole mass and the absolute magnitude of the galaxy in different bands. As the correlation is based on data from nearby galaxies, we use k+e corrected absolute magnitudes, the k+e corrections are from Bicker et al. (\cite{Bi04}).

The correlations used are: Gebhardt et al. (\cite{Ge03}) (absolute B-band magnitude), Bettoni et al. (\cite{Be03}) and McLure \& Dunlop (\cite{MD01}) (absolute R-band magnitude) and Marconi \& Hunt (\cite{MH03}) (absolute J-band magnitude).

The determined values are upper limits, as the k+e-corrected values were used, they describe the mass of the central black hole if the galaxy evolves until $z=0$. In addition, the central black hole could have gathered mass by an unknown amount. The results for each method and the averaged value can be found in Tab. \ref{MBH0809}.

Note that the host galaxy of FDF0809 was clearly identified as spiral galaxy but the correlations are based on data from early-type galaxies. Pastorini et al. (\cite{Pa07}) found recently that nearby spiral galaxies seem to follow the $M_{BH}-L_{bulge}$ correlations based on early-type galaxies. Though the sample used in this paper is extremely small (6 values for $M_{BH}$ and 2 upper limits). As we do not know the bulge--disk ratio for the host galaxy of FDF0809, we are overestimating the bulge luminosity and thus the black hole mass. Additionally, following errors have to be taken into account: systematic errors between different models, errors in the values of the apparent magnitude and the k+e-corrections and errors determining the black hole masses for the nearby galaxies the relations are based on. All in all this yields a rough estimate for the upper limit of the black hole mass.

\begin{table*}
\caption{Mass of central black hole for FDF0809 using different methods.}
\centering
\begin{tabular}[h]{ccc}
\hline \hline
Ref. & Filter & $log(M_{BH}) [M_{\odot}]$\\
\hline 
Gebhardt & B & $<$8.48 \\
Marconi\&Hunt & J & $<$7.39 \\
Bettoni & R & $<$8.39 \\
McLure\&Dunlop & R & $<$8.29 \\
average & - & $<$8.1 \\ 
\hline
\end{tabular}
\label{MBH0809}
\end{table*}

\section{Comparison with previous work and discussion}

In this chapter we will compare our results to those of other authors and discuss the results regarding the link between galaxy evolution and quasar activity and the evolution of supermassive black holes.

The apparent co-evolution of quasar activity and star formation rate density (Nandra et al. \cite{Na05}, Appenzeller et al. \cite{Ap04}) leads to the question if quasar activity and heavy star formation are directly linked. I.e. are both phenomena triggered by the same event or does one of the events trigger the other.

Hyv\"{o}nen et al. (\cite{Hy07}) studied low redshift ($z<0.3$) BL Lac host galaxies and found typical elliptical galaxies with an additional young stellar populations. They did not find signs of intercation. They also analysed restframe B-R colours for their low-redshift BL Lac objects and found values in the range of $0.5$--$2$ mag. Restframe B--R corresponds to I--J for FDF0809, giving a value of $0.54$ mag, this is well consistent with the findings of Hyv\"{o}nen at al. (\cite{Hy07}).

Sanchez et al. (\cite{Sa04}) studied low redshift ($0.5<z<1.1$) AGN host galaxies and found late as well as early type morphologies. Most of the galaxies in their sample showed signs of interaction, young stellar populations or even ongoing star formation.

Hutchings et al. (\cite{Hu02}) studied high redshift ($z\sim2$) host galaxies. They found early type morphologies with signs of interaction and ongoing star formation.

Hutchings (\cite{Hu03}) resolved the host galaxies of 4 high redshift quasars ($z\sim4.7$) and found heavily disturbed galaxies, partially showing signs of ongoing merging.

Jahnke et al. (\cite{Ja04}) studied host galaxies of moderately luminosity quasars in the rest frame UV at high redshifts ($1.8<z<2.75$). With a mean SFR of $\sim12M_{\odot}/yr$ (K. Jahnke, private communication), the galaxies are clearly star forming. The galaxies show magnitudes at restframe $200nm$ in the range --$20<M_{200nm}<$--$22.2$. A rest frame wavelength of $200nm$ corresponds to R band for FDF6007. Depending on the galaxy model used for the fits we measure $M=$--$21.75/$--$22.15$, thus our values are well consistent with those of Jahnke at al. (\cite{Ja04}). Comparing the luminosities to those of the two possibly resolved host galaxies, FDF1837 ($M=$--$21.09/$--$21.66)$ in observed frame R) is well consistent with the values found be Jahnke et al. (\cite{Ja04}), whereas FDF2229 ($M=$--$23.14/$--$23.11$ in observed frame R) is slightly more luminous.

Our resolved host galaxies thus seem well consistent with the host galaxies analysed by other authors. For a general trend it seems that almost all quasar host galaxies show either ongoing star formation, signs of interaction or stellar populations that imply recent massive star formation. This implies that quasar activity and and star formation are indeed linked. However they might not happen simultaneously. It also implies that the assumed passive evolution of quasar host galaxies (see e.g. Falomo et al. \cite{Fa05}) that is based on samples of host galaxies that have been observed in one band only might not hold. For most of these studies, the authors assumed old stellar populations, typical for massive ellipticals, which seems contradictory to the findings from multi-band data.

Another open question is the formation of supermassive black holes. It is still not clear when and how they formed. Measuring black hole masses for high redshift quasars could give hints when supermassive black holes formed. All samples so far are heavily flux limited (e.g. Vestergaard et al. \cite{Ve08}, data from SDSS). Thus measuring black hole masses in high redshift low luminosity quasars might give a hint on when black holes formed.

Kuhlbrodt et al. (\cite{Ku05}) measured the black hole masses of five high redshift ($z\sim2.2$) high luminosity quasars and found black hole masses of $\sim10^{9}M_{\odot}$. With black hole masses in the range $7.04<log(M_{BH})<8.94$, all values from our sample are lower than the ones found by Kuhlbrodt et al. (\cite{Ku05}). However, this is sensible as we studied low to moderate luminosity quasars whereas Kuhlbrodt et al. (\cite{Ku05}) studied high luminosity quasars.

\section{Summary and Conclusions}

In this paper we have presented a multiband (eight filters from near UV to NIR) analysis of eight high-redshift quasars with redshifts between $z=0.87$ and $z=3.37$. For two of the quasars (FDF0809 at $z=0.87$ and FDF6007 at $z=2.75$) the host galaxies could be resolved. For two quasars the host galaxy is regarded as marginally resolved, therefore magnitudes found are merely upper limits. For the resolved and marginally resolved host galaxies we analysed their age, stellar populations and SFR and estimated their central black hole masses. The results can be summarised as follows:

By analysing the colours of the resolved host galaxies we tried to constrain the galaxy type and formation redshift of the galaxy. FDF0809 is consistent with an either E/S0 or a Sb-type galaxy, while extreme bursts can clearly be rejected. Additionally, using simulation of spectra we were able to confirm stellar populations consistent with a spiral type galaxy. For the other three host galaxies, the colours are consistent with either an E/S0 or Sb-type galaxy with a formation at $z=5$ or $z=3$ or a long burst with a formation at $z=3$.

The determination of the star formation rate is critical, as the used model breaks down if the star formation rate changes on small time scales ($\lesssim10^{8}$yr). This limitation of the model was verified by calculating SFRs for model galaxies. Thus the results should be handled with care, as the SFR is potentially overestimated. For all objects, the SFR is estimated to be between 0.03 and 33 $M_{\odot}/yr$, which is actually consistent with values derived by Jahnke et al. (\cite{Ja04}) using the same method at similar redshifts.

The mass of the central black hole was calculated using the C IV line width for 6/8 sources. Within the uncertainties we derived BH masses between $\sim 10^{7}$ and $10^{9} M_{\odot}$. The results are consistent with those of Vestergaard et al. (\cite{Ve08}) and Kuhlbrodt et al. (\cite{Ku05}), considering the fact that both studies observed quasars more luminous than those from our sample.

Both clearly resolved host galaxies are consistent with those studied by other authors. Being a late-type galaxy, FDF0809 fits well into the picture of quasar host galaxies at low redshift being hosted by low luminosity quasars. FDF6007, showing moderate star formation rate and a UV luminosity consistent with those of similar host galaxies studied by Jahnke at al. (\cite{Ja04}), seems to be no expection either. Both fit into the general picture of quasar host galaxies being either late-type galaxies, like FDF0809, interacting galaxies or galaxy that have recently undergone strong star formation (FDF6007 might fit into this class).

Analysis of high-redshift quasar hosts on the one hand requires extremely deep images to guarantee high signal-to-noise. Otherwise the fainter host galaxies can not be detected. On the other hand high resolution imaging is necessary to be able to resolve high redshift hosts, which are just merely more extended than a point-like source. Further exploration of the evolution of the mass of the central black hole and properties of the galaxies like SFR or stellar populations will help to understand triggering and evolution of quasars, their host galaxies and galaxies in general. 

\begin{acknowledgements}
We acknowledge EC funding under contract HPRN-CT-2002-00321 (ENIGMA). This work has been supported by the Deutsche Forschungsgemeinschaft (SFB 375, SFB 439) and the German Federal Ministry of Science and Technology (Grants 05 2HD50A, 05 2GO20A and 05 2MU104). We thank Knud Jahnke and Tapio Pursimo for helpfull and interesting discussions and Maurilio Pannella for providing the PSF for the HST data. We also thank the anonymous referee for his/her critical but constructive comments.
\end{acknowledgements}

\newpage

\appendix
\section{Result tables}

The meanings of the columns are as follows:
\begin{itemize}
 \item $\chi^{2}_{c}$: $\chi^{2}$ for core-fit
 \item $\chi^{2}_{g}$: $\chi^{2}$ for core+galaxy-fit
 \item $m_{c}$: magnitude of central point source for each fit respectively
 \item $M_{c}$: absolute magnitude of central point source for each fit respectively
 \item $m_{g}$: magnitude of galaxy for each fit respectively
 \item $M_{g,k}$: magnitude of galaxy, k-corrected
 \item $M_{g,e}$: magnitude of galaxy, k- and evolution-corrected
 \item $k$: k-correction
 \item $e$: evolution-correction 
 \item $r_{e}$['']: effective radius of galaxy in arcsec
 \item $r_{e}$[kpc]: effective radius of galaxy in kpc
 \item flag: results from fits
\begin{itemize}
	\item b: resolved, bulge
	\item d: resolved, disk
	\item bd: resolved, either bulge or disk
	\item u: unresolved
\end{itemize}
 \item resolved?: results from error simulations:
\begin{itemize}
	\item $3\sigma$: error simulations yielded resolution with $3\sigma$-significance
	\item $5\sigma$: error simulations yielded resolution with $5\sigma$-significance
	\item x: not resolved according to error simulations
	\item u: unresolved in fits, no error simulations done
\end{itemize}
\end{itemize}
If a value is marked as ''--'', the object was not resolved or could not be fitted due to the object being too faint, in these cases values for the core can be found in the table for the unresolved objects. For the NIR-filters J and Ks, only the fits for FDF0809 yielded reasonable results, thus they are not mentioned for the other objects. K- and evolution corrections from Bicker et al. (\cite{Bi04}) were applied for FDF0809 only. Thus for the absolute magnitudes for the other objects, filters listed do not correspond to rest-frame filters but filters transfered to the observing frame of the related object. Cosmology is $H_{0}=70\textrm{km/s/Mpc}, \Omega_{\Lambda}=0.7, \Omega_{m}=0.3$, unless for evolution-corrected magnitudes of the host galaxy of FDF0809, where cosmology is: $H_{0}=65\textrm{km/s/Mpc}, \Omega_{0}=0.1$..

\begin{table*}
\caption{Results for magnitude of core for the unresolved objects (results from core-fit). If a value is marked as --, the object was resolved, and the values can be found in the table for the resolved objects, if a value is marked as x, it was not detected in this band.}
\centering
\begin{tabular}{cccccccc}
\hline \hline
Filter & FDF1837 & FDF2229 & FDF2633 & FDF4683 & FDF5962 & FDF6007 & FDF6233 \\
 & $m_{c}$ / $M_{c}$ & $m_{c}$ / $M_{c}$ & $m_{c}$ / $M_{c}$ & $m_{c}$ / $M_{c}$ & $m_{c}$ / $M_{c}$ & $m_{c}$ / $M_{c}$ & $m_{c}$ / $M_{c}$ \\
\hline
U & -- & -- & 24.41 & 21.52 & 22.53 & 26.48 & x \\
 & -- & -- & --22.68 & --25.81 & --23.07 & --20.32 & x \\
B & -- & -- & 24.45 & 20.14 & 22.70 & -- & 26.04 \\
 & -- & -- & --22.64 & --27.19 & --22.90 & -- & --20.31 \\
g & -- & -- & 23.55 & 19.35 & 22.97 & -- & 25.36 \\
 & -- & -- & --23.54 & --27.98 & --22.63 & -- & --20.99 \\
R & -- & -- & 23.02 & 18.90 & 22.58 & -- & 24.11 \\
 & -- & -- & --24.07 & --28.43 & --23.02 & -- & --22.24 \\
I & -- & -- & 22.69 & 18.55 & 21.89 & -- & 23.84 \\
 & -- & -- & --24.40 & --28.78 & --23.71 & -- & --22.51 \\
J & 23.80 & 21.69 & 24.01 & 19.51 & 23.43 & x & 23.72 \\
 & --22.59 & --24.47 & --23.08 & --27.82 & --22.17 & x & --22.63 \\
Ks & 21.59 & 19.73 & 22.23 & 17.52 & 21.10 & 22.25 & 21.78 \\
 & --24.80 & --26.43 & --24.86 & --29.81 & --24.50 & --24.55 & --24.57 \\
F814W & 22.98 & 20.61 & 22.76 & 18.71 & 21.74 & 25.01 & 23.80 \\
 & --23.41 & --25.55 & --24.33 & --28.62 & --23.86 & --21.79 & --22.55 \\
\hline
\end{tabular}
\end{table*}

\begin{sidewaystable*}
\caption{Results for FDF0809, values are from the fit with variable ellipticity and PA, apart from filters J, Ks and F814W, where no fit with variable ellipticity/PA is available.}
\centering
\begin{tabular}{cccccccccccccc}
\hline \hline
filter & $\chi^{2}_{c}$ & $\chi^{2}_{g}$ & $m_{c}$ & $M_{c}$ & $m_{g}$ & $k$ & $e$ & $M_{g,k}$ & $M_{g,e}$ & $r_{e}$ & $r_{e}$ & flag & resolved? \\
(fit) & & & & & & & & & & [''] & [kpc] & & \\
\hline
U(b) & 1.656 & 1.146 & 22.52$\pm$0.07 & --21.20 & 23.08$\pm$0.15 & 3.5 & --3.1 & --24.12 & --21.03 & 0.19$\pm$0.04 & 1.5 & bd & 3$\sigma$ \\
U(d) & 1.656 & 1.102 & 22.77$\pm$0.06 & --20.95 & 22.83$\pm$0.07 & 0.1 & --1.1 & --21.02 & --19.94 & 0.23$\pm$0.01 & 1.8 & bd & 5$\sigma$ \\ 
B(b) & 4.321 & 2.372 & 22.85$\pm$0.08 & --20.87 & 24.63$\pm$0.42 & 3.4 & --2.2 & --22.52 & --20.06 & 0.21$\pm$0.07 & 1.6 & d & 3$\sigma$ \\
B(d) & 4.321 & 2.083 & 22.87$\pm$0.08 & --20.85 & 24.62$\pm$0.23 & 0.9 & --1.0 & --20.02 & --19.04 & 0.24$\pm$0.06 & 1.9 & d & 3$\sigma$ \\
g(b) & 3.173 & 1.525 & 22.90$\pm$0.12 & --20.82 & 23.62$\pm$0.31 & 2.4 & --1.4 & --22.52 & --21.13 & 0.20$\pm$0.06 & 1.5 & bd & 3$\sigma$ \\
g(d) & 3.173 & 1.422 & 22.99$\pm$0.43 & --20.73 & 23.54$\pm$0.32 & 1.1 & --0.9 & --21.32 & --20.44 & 0.24$\pm$0.05 & 1.9 & bd & 3$\sigma$ \\
R(b) & 25.470 & 1.700 & 23.01$\pm$0.47 & --20.71 & 22.25$\pm$0.27 & 1.7 & --1.2 & --23.12 & --21.93 & 0.26$\pm$0.22 & 2.0 & bd & x \\
R(d) & 23.470 & 1.823 & 22.51$\pm$0.06 & --21.21 & 22.75$\pm$0.16 & 1.0 & --0.8 & --22.02 & --21.24 & 0.41$\pm$0.07 & 3.2 & bd & 5$\sigma$ \\
I(b) & 114.999 & 6.421 & 23.11$\pm$0.96 & --20.61 & 21.43$\pm$0.17 & 0.8 & --1.0 & --23.12 & --22.13 & 0.26$\pm$0.13 & 2.0 & d & x \\
I(d) & 114.999 & 2.472 & 22.47$\pm$0.10 & --21.25 & 21.79$\pm$0.09 & 0.3 & --0.7 & --22.22 & --21.54 & 0.32$\pm$0.02 & 2.5 & d & 5$\sigma$ \\
F814W(b) & 59.105 & 6.919 & 22.33$\pm$0.17 & --21.39 & 21.28$\pm$0.10 & 0.9 & --1.1 & --23.34 & --22.25 & 0.51$\pm$0.27 & 3.9 & bd & x \\
F814W(d) & 59.105 & 8.169 & 22.15$\pm$0.09 & --21.57 & 21.86$\pm$0.04 & 0.4 & --0.7 & --22.26 & --21.58 & 0.30$\pm$0.03 & 2.3 & bd & 5$\sigma$ \\
J(b) & 4.159 & 1.846 & 27.21$\pm$1.09 & --16.51 & 20.97$\pm$0.07 & 0.1 & --0.9 & --22.62 & --21.73 & 0.27$\pm$0.03 & 2.1 & d & 5$\sigma$ \\
J(d) & 4.159 & 1.702 & 23.04$\pm$3.3 & --20.68 & 21.25$\pm$0.16 & --0.1 & --0.4 & --22.62 & --22.24 & 0.40$\pm$0.05 & 3.1 & d & 5$\sigma$ \\
Ks(b) & 3.291 & 1.904 & 21.19$\pm$1.61 & --22.53 & 19.58$\pm$0.30 & --0.8 & --0.8 & --23.32 & --22.53 & 0.30$\pm$0.09 & 2.3 & d & 3$\sigma$ \\
Ks(d) & 3.291 & 1.312 & 23.68$\pm$2.18 & --20.04 & 19.35$\pm$0.10 & --0.8 & --0.2 & --23.62 & --23.44 & 0.45$\pm$0.07 & 3.5 & d & 5$\sigma$ \\
\hline
\end{tabular}
\end{sidewaystable*}

\begin{table*}
\caption{Results for FDF1837.}
\centering
\begin{tabular}{cccccccccccccc}
\hline
filter & $\chi^{2}_{c}$ & $\chi^{2}_{g}$ & $m_{c}$ & $M_{c}$ & $m_{g}$ & $M_{g}$ & $r_{e}$ & $r_{e}$ & flag & resolved? \\
(fit) & & & & & & & [''] & [kpc] & & \\
\hline
\hline
1837 & & & & & & & & & & & & & \\
U(b) & 1.168 & 0.912 & 24.46$\pm$0.17 & --21.93 & 24.28$\pm$0.19 & --21.99 & 0.38$\pm$0.17 & 3.1 & bd & x \\
U(d) & 1.168 & 0.905 & 24.44$\pm$0.21 & --21.95 & 24.48$\pm$0.22 & --21.79 & 0.32$\pm$0.17 & 2.6 & bd & x \\ 
B(b) & 2.046 & 1.859 & 24.02$\pm$0.05 & --22.37 & 26.32$\pm$0.41 & --19.95 & 0.30$\pm$0.08 & 2.5 & bd & 3$\sigma$ \\
B(d) & 2.046 & 1.819 & 24.02$\pm$0.07 & --22.37 & 26.49$\pm$0.90 & --19.78 & 0.26$\pm$0.03 & 2.1 & bd & 5$\sigma$ \\
g(b) & 2.306 & 1.410 & 24.36$\pm$0.11 & --22.03 & 24.58$\pm$0.19 & --19.69 & 0.31$\pm$0.11 & 2.6 & bd & x \\
g(d) & 2.306 & 1.365 & 24.24$\pm$0.15 & --22.15 & 24.91$\pm$0.38 & --21.36 & 0.38$\pm$0.19 & 3.1 & bd & x \\ 
R(b) & 1.576 & 1.376 & 23.90$\pm$0.06 & --22.49 & 25.18$\pm$0.99 & --21.09 & 0.14$\pm$0.09 & 1.6 & d & x \\
R(d) & 1.576 & 1.267 & 24.18$\pm$0.15 & --22.21 & 24.61$\pm$0.65 & --21.66 & 0.13$\pm$0.07 & 1.1 & d & x \\ 
I(b) & 3.349 & 2.053 & 23.31$\pm$0.06 & --23.08 & 24.04$\pm$0.19 & --22.23 & 0.11$\pm$0.06 & 0.9 & d & x \\
I(d) & 3.349 & 1.583 & 23.53$\pm$0.59 & --22.86 & 23.79$\pm$0.48 & --22.48 & 0.12$\pm$0.04 & 1.0 & d & 3$\sigma$ \\
F814W & -- & -- & -- & -- & -- & -- & -- & -- & u & u \\
\hline
\end{tabular}
\end{table*}

\begin{table*}
\caption{Results for FDF2229.}
\centering
\begin{tabular}{cccccccccccccc}
\hline \hline
filter & $\chi^{2}_{c}$ & $\chi^{2}_{g}$ & $m_{c}$ & $M_{c}$ & $m_{g}$ & $M_{g}$ & $r_{e}$ & $r_{e}$ & flag & resolved? \\
(fit) & & & & & & & [''] & [kpc] & & \\
\hline
U(b) & 3.738 & 2.637 & 21.48$\pm$0.05 & --24.68 & 22.76$\pm$0.18 & --23.40 & 0.21$\pm$0.06 & 1.7 & bd & 3$\sigma$ \\
U(d) & 3.738 & 2.645 & 21.58$\pm$0.16 & --24.58 & 22.53$\pm$0.23 & --23.63 & 0.20$\pm$0.05 & 1.7 & bd & 3$\sigma$ \\
B(b) & 9.696 & 3.308 & 22.16$\pm$0.10 & --24.00 & 23.81$\pm$0.67 & --22.35 & 0.31$\pm$0.08 & 2.6 & bd & 3$\sigma$ \\
B(d) & 9.696 & 3.595 & 22.11$\pm$0.11 & --24.05 & 24.28$\pm$0.54 & --21.88 & 0.38$\pm$0.09 & 3.2 & bd & 3$\sigma$ \\
g(b) & 5.872 & 3.214 & 21.87$\pm$0.16 & --24.29 & 23.54$\pm$0.73 & --22.62 & 0.28$\pm$0.09 & 2.3 & bd & 3$\sigma$ \\
g(d) & 5.872 & 3.339 & 21.80$\pm$0.05 & --24.36 & 24.10$\pm$0.28 & --22.06 & 0.47$\pm$0.05 & 3.9 & bd & 5$\sigma$ \\
R(b) & 7.808 & 4.179 & 21.29$\pm$0.06 & --24.87 & 23.02$\pm$0.67 & --23.14 & 0.16$\pm$0.09 & 1.3 & d & x \\
R(d) & 7.808 & 3.873 & 21.30$\pm$0.05 & --24.86 & 23.05$\pm$0.62 & --23.11 & 0.20$\pm$0.07 & 1.7 & d & x \\ 
I(b) & 11.985 & 9.452 & 20.84$\pm$0.03 & --25.32 & 23.07$\pm$0.83 & --23.09 & 0.13$\pm$0.06 & 1.1 & d & x \\
I(d) & 11.985 & 8.589 & 20.89$\pm$0.36 & --25.27 & 22.79$\pm$1.05 & --23.37 & 0.13$\pm$0.20 & 1.1 & d & x \\ 
F814W & -- & -- & -- & -- & -- & -- & -- & -- & u & u \\
\hline
\end{tabular}
\end{table*}

\begin{table*}
\caption{Results for FDF6007.}
\centering
\begin{tabular}{cccccccccccccc}
\hline \hline
filter & $\chi^{2}_{c}$ & $\chi^{2}_{g}$ & $m_{c}$ & $M_{c}$ & $m_{g}$ & $M_{g}$ & $r_{e}$ & $r_{e}$ & flag & resolved? \\
(fit) & & & & & & & [''] & [kpc] & & \\
\hline
U & -- & -- & -- & -- & -- & -- & -- & -- & u & u \\
B(b) & 3.778 & 1.271 & 26.61$\pm$0.24 & --20.19 & 25.20$\pm$0.16 & --21.60 & 0.5$\pm$0.19 & 3.9 & bd & x \\
B(d) & 3.778 & 1.246 & 26.19$\pm$0.16 & --20.61 & 25.70$\pm$0.21 & --21.10 & 0.44$\pm$0.13 & 3.5 & bd & 3$\sigma$ \\
g(b) & 1.244 & 1.136 & 26.39$\pm$0.42 & --20.41 & 25.99$\pm$0.58 & --20.81 & 0.24$\pm$0.16 & 1.9 & bd & x \\
g(d) & 1.244 & 1.130 & 26.42$\pm$0.67 & --20.38 & 26.02$\pm$0.33 & --20.78 & 0.25$\pm$0.12 & 2.0 & bd & x \\ 
R(b) & 1.685 & 1.034 & 30.74$\pm$0.16 & --16.06 & 24.65$\pm$0.07 & --22.15 & 0.12$\pm$0.02 & 1.0 & bd & 5$\sigma$ \\
R(d) & 1.685 & 1.047 & 26.15$\pm$0.90 & --20.65 & 25.06$\pm$0.29 & --21.74 & 0.23$\pm$0.13 & 1.8 & bd & x \\ 
I(b) & 2.080 & 1.045 & 25.20$\pm$0.49 & --21.78 & 24.49$\pm$0.12 & --22.31 & 0.67$\pm$0.15 & 5.3 & bd & 3$\sigma$ \\
I(d) & 2.080 & 1.054 & 24.93$\pm$0.23 & --21.87 & 25.04$\pm$0.10 & --21.76 & 0.54$\pm$0.08 & 4.3 & bd & 5$\sigma$ \\
F814W & -- & -- & -- & -- & -- & -- & -- & -- & u & u \\
\hline
\end{tabular}
\end{table*}

\end{document}